\documentclass{amsart}

\usepackage{psfig}
\usepackage{latexsym}
\usepackage{amssymb}
\usepackage{mathrsfs}
\usepackage{setspace}
\numberwithin{equation}{section}

\theoremstyle{plain}
\newtheorem{thm}{Theorem}[section]
\newtheorem{prop}{Proposition}[section]
\newtheorem{lemma}{Lemma}[section]

\theoremstyle{definition}
\newtheorem{definition}{Definition}[section]
\renewcommand{\d}{\ensuremath{\partial}}
\newcommand{\dt}{\ensuremath{\partial_{\tau}}}

\newcommand{\da}{\ensuremath{D^{\alpha}}}

\newcommand{\dae}{\ensuremath{D^{\alpha_{1}}}}
\newcommand{\dat}{\ensuremath{D^{\alpha_{2}}}}
\newcommand{\dah}{\ensuremath{D^{\hat{\alpha}}}}
\newcommand{\ett}{\ensuremath{e^{-2\tau}}}
\renewcommand{\t}{\ensuremath{{}_{\tau}}}
\renewcommand{\tt}{\ensuremath{{}_{\tau\tau}}}

\newcommand{\vre}{\ensuremath{\mathcal{E}}}
\newcommand{\vrf}{\ensuremath{\mathcal{F}}}
\newcommand{\tvre}{\ensuremath{\tilde{\mathcal{E}}}}

\newcommand{\et}{\ensuremath{\tilde{E}}}
\newcommand{\ft}{\ensuremath{\tilde{f}}}

\newcommand{\m}{\ensuremath{\mathrm{max}}}
\newcommand{\n}{{}_{\ensuremath{n}}}
\newcommand{\npo}{{}_{\ensuremath{n+1}}}
\newcommand{\nmo}{{}_{\ensuremath{n-1}}}
\newcommand{\nc}{{}_{\ensuremath{n,}}}
\newcommand{\npoc}{{}_{\ensuremath{n+1,}}}
\newcommand{\nmoc}{{}_{\ensuremath{n-1,}}}
\newcommand{\qt}{\ensuremath{\tilde{Q}}}
\newcommand{\pt}{\ensuremath{\tilde{P}}}
\newcommand{\at}{\ensuremath{\tilde{\alpha}}}
\newcommand{\ah}{\ensuremath{\hat{\alpha}}}

\newcommand{\e}{\ensuremath{\epsilon}}
\renewcommand{\a}{\ensuremath{\alpha}}
\renewcommand{\b}{\ensuremath{\beta}}
\newcommand{\de}{\ensuremath{\delta}}
\newcommand{\g}{\ensuremath{\gamma}}
\renewcommand{\ae}{\ensuremath{\alpha_{1}}}
\renewcommand{\at}{\ensuremath{\alpha_{2}}}
\newcommand{\vpt}{\ensuremath{\tilde{\mathcal{P}}}}

\newcommand{\vqt}{\ensuremath{\tilde{\mathcal{Q}}}}

\newcommand{\bP}{\mathbf{P}}

\newcommand{\bp}{\mathbf{p}}
\newcommand{\bF}{\mathbf{F}}
\newcommand{\ma}{\mathrm{max}}
\newcommand{\mi}{\mathrm{min}}

\begin{document}
\title{On Gowdy vacuum spacetimes}
\author{Hans Ringstr\"{o}m}
\address{Max-Planck-Institut f\"{u}r Gravitationsphysik,
 Am M\"{u}hlenberg 1,D-14476 Golm, Germany}

%

\begin{abstract}
By Fuchsian techniques, a large family of Gowdy vacuum spacetimes 
have been constructed for which one has detailed control over the
asymptotic behaviour. In this paper we formulate a condition on
initial data yielding the same form of asymptotics. 
\end{abstract}
\maketitle

\section{Introduction}

This paper is concerned with the study of 
cosmological singularities. By a cosmological spacetime we mean
a globally hyperbolic Lorentz manifold with compact spatial Cauchy
surfaces satisfying Einstein's equations. A singularity is
characterized by causal geodesic incompleteness (assuming the
spacetime satisfies some natural maximality condition). Causal geodesic
incompleteness, and thus singularities, is guaranteed in general 
situations by the singularity theorems. However, the question of 
curvature blow up at the singularity, and the related question of 
strong cosmic censorship are a separate issue. The desire to answer
these questions motivated this paper.

Most of the work in the area of cosmological singularities has
concerned the spatially homogeneous case. However,
some classes of spatially inhomogeneous spacetimes have been studied
analytically and numerically. In particular, the so called Gowdy
spacetimes have received considerable attention. The reason for this
is probably the fact that analyzing the Gowdy spacetimes is on the borderline
of what is doable and what is not. These spacetimes were first 
introduced in \cite{gowdy} (see also \cite{chr1}), and in \cite{mon}
the basic questions concerning global
existence were answered. We will take the Gowdy vacuum spacetimes on
$\mathbb{R}\times T^{3}$ to be metrics of the form (\ref{eq:gowdy}). 
However, some sort of motivation for this choice seems to be in order.
Below, we give a rough description of more natural conditions that
lead to this form of metric. In fact, the conditions below do not
imply the form (\ref{eq:gowdy}), see \cite{chr1} pp. 116-117.
However, the discrepancy can be eliminated by a coordinate
transformation which is local in space. Combining this observation
with domain of dependence arguments hopefully convinces the reader 
that nothing essential is lost by considering metrics of the form
(\ref{eq:gowdy}). The description below is brief and we  refer the
interested reader to \cite{gowdy} and \cite{chr1} for more details. 
The following conditions can be used to define the Gowdy 
vacuum spacetimes:
\begin{itemize}
\item It is an orientable globally hyperbolic vacuum spacetime.
\item It has compact spatial Cauchy surfaces.
\item There is a smooth effective group action of $U(1)\times U(1)$ on the
      Cauchy surfaces under which the metric is invariant.
\item The twist constants vanish.
\end{itemize}
Let us explain the terminology.
A group action of a Lie group $G$ on a manifold $M$ is effective if
$gp=p$ for all $p\in M$ implies $g=e$. Due to the existence of the
symmetries we get two Killing fields. Let us call them $X$ and $Y$.
The twist constants are defined by
\[
\kappa_{X}=\e_{\a\b\g\de}X^{\a}Y^{\b}\nabla^{\g}X^{\de}\ \ \
\mathrm{and}\ \ \
\kappa_{Y}=\e_{\a\b\g\de}X^{\a}Y^{\b}\nabla^{\g}Y^{\de}.
\]
The fact that they are constants is due to the field equations. 
By the existence of the effective group action, one can draw the
conclusion that the spatial Cauchy surfaces have topology $T^{3}$,
$S^{3}$, $S^{2}\times S^{1}$ or a Lens space. 
In all the cases except
$T^{3}$, the twist constants have to vanish. However, in the case of 
$T^{3}$ this need not be true, and the condition that they vanish is
the most unnatural of the ones on the list above. There is however a
reason for separating the two cases. Considering the case of $T^{3}$
spatial Cauchy surfaces, numerical studies indicate that the Gowdy
case is convergent \cite{bag} and the general case is oscillatory
\cite{baiw}. Analytically analyzing the case with non-zero twist
constants can therefore reasonably be expected to be significantly
more difficult than the Gowdy case. In this paper we will consider
the Gowdy $T^{3}$ case.

Due to the numerical studies, cf. \cite{bag}, the picture as to what
should happen is quite clear. In order to 
formulate the conclusions, we need to parametrize the metric. One
way of doing so is
\begin{equation}\label{eq:gowdy}
g=e^{(\tau-\lambda)/2}(-e^{-2\tau}d\tau^{2}+d\theta^2)
+e^{-\tau}[e^{P}d\sigma^2+2e^{P}Qd\sigma d\delta+
(e^{P}Q^2+e^{-P})d\delta^2].
\end{equation}
Here, $\tau\in\mathbb{R}$ and $(\theta,\sigma,\delta)$ are coordinates
on $T^{3}$. The evolution equations become
\begin{eqnarray}
P_{\tau\tau}-e^{-2\tau}P_{\theta\theta}-
e^{2P}(Q_{\tau}^2-e^{-2\tau}Q_{\theta}^2) & = & 0 \label{eq:ge1}\\
Q_{\tau\tau}-e^{-2\tau}Q_{\theta\theta}
+2(P_{\tau}Q_{\tau}-e^{-2\tau}P_{\theta}Q_{\theta}) & = & 0,\label{eq:ge2}
\end{eqnarray}
and the constraints
\begin{eqnarray}
\lambda_{\tau} & = & P_{\tau}^{2}+e^{-2\tau}P_{\theta}^{2}+
e^{2P}(Q_{\tau}^{2}+e^{-2\tau}Q_{\theta}^{2})\label{eq:gc1}\\
\lambda_{\theta} & = & 2(P_{\theta}P_{\tau}+e^{2P}Q_{\theta}Q_{\tau}).
\label{eq:gc2}
\end{eqnarray}
Obviously, the constraints are decoupled from the evolution equations, 
excepting the condition on $P$ and $Q$ implied by (\ref{eq:gc2}).
Thus the equations of interest are the two non-linear coupled wave
equations (\ref{eq:ge1})-(\ref{eq:ge2}). In this parametrization, the
singularity corresponds to $\tau\rightarrow \infty$, and the subject
of this article is the asymptotics of solutions to 
(\ref{eq:ge1})-(\ref{eq:ge2}) as $\tau\rightarrow \infty$. 
The asymptotics we derive will then be used to obtain conclusions
concerning curvature blow up. There is a special case of
these equations determined by the condition $Q=0$ which is called the
polarized case. This has been handled in \cite{jam}, which also
considers the other topologies. The asymptotic behaviour of the
solution $P$ to (\ref{eq:ge1}) in the polarized case is given by
\[
P(\tau,\theta)=v(\theta)\tau+\phi(\theta)+e^{-\e\tau}u(\tau,\theta)
\]
where $\epsilon>0$ and $u(\tau,\theta)\rightarrow 0$ as
$\tau\rightarrow 0$. In this
situation, $v$ and $\phi$ are arbitrary smooth functions on the
circle. In the general case, the numerical studies indicate that 
the ``velocity'' $v$ should typically be confined to the open interval
$(0,1)$. To be more precise, the following asymptotics are expected
in general:
\begin{eqnarray}
P(\tau,\theta) & = & v(\theta)\tau+\phi(\theta)+e^{-\epsilon\tau}
u(\theta,\tau)\label{eq:as1}\\
Q(\tau,\theta) & = & q(\theta)+e^{-2v(\theta)\tau}[\psi(\theta)+
w(\tau,\theta)]\label{eq:as2}
\end{eqnarray}
where $\epsilon>0$ and $w,u\rightarrow 0$ as $\tau\rightarrow \infty$ and
$0<v(\theta)<1$. A heuristic argument motivating the condition on 
the velocity can be found in \cite{bag}. However, the numerical
simulations also indicate the occurrences of ``spikes''. Let us describe
one sort of spike that can occur. It can happen that at a
spatial point $\theta_{0}$, $P_{\tau}$ will, in the limit, have a 
value greater than $1$ whereas the limiting values of $P_{\tau}$ in
a punctured neighbourhood of $\theta_{0}$, $P_{\tau}$ belong to $(0,1)$. 
Furthermore, $Q$ converges nicely in a neighbourhood of $\theta_{0}$ 
with a zero of the spatial derivative of $Q$ at
$\theta_{0}$. Beyond the numerical indications of these types of 
features, a family of solutions with spikes have been constructed
in \cite{raw}, so that the behaviour described above is known to 
occur. The type of spike described above is a ``true'' spike. There 
are also other types of spikes called ``false'' spikes at which
$Q$ has a discontinuity. We refer the reader to \cite{raw} for more
details. One relevant question to ask is whether the spikes are a 
result of a bad parametrization of the metric or if they really have
a geometrical significance. It seems that the false spikes are a 
result of bad parametrization, but the true spikes can be detected by
curvature invariants \cite{raw}. 

In this paper we will not be concerned with spikes, but will focus on
solutions with an asymptotic behaviour of the form 
(\ref{eq:as1})-(\ref{eq:as2}). By the so called Fuchsian
techniques one can construct a large family of solutions
with such asymptotic behaviour. In fact, given functions
$v,\phi,q$ and $\psi$ from $S^{1}$ to $\mathbb{R}$ of a suitable
degree of smoothness and subject to the condition $0<v<1$, one can
construct solutions to (\ref{eq:ge1})-(\ref{eq:ge2}) with asymptotics
of the form (\ref{eq:as1})-(\ref{eq:as2}). The proof of this in the
real analytic case can be found in \cite{kar1} and \cite{r} covers
the smooth case. One nice feature of this construction is the fact
that one gets to specify four functions freely, just as as if though
one were specifying initial data for (\ref{eq:ge1})-(\ref{eq:ge2}).

The purpose of this paper is to provide a condition on the initial
data yielding the asymptotic behaviour (\ref{eq:as1})-(\ref{eq:as2}).
There are several reasons for wanting to prove such a statement. As
was mentioned above, one can construct a large family of solutions
with the desired asymptotic behaviour, but it is not clear how big
this family is in terms of initial data. In this paper we prove the
existence of an open set of initial data yielding the desired
asymptotics. Observe that the equations (\ref{eq:ge1})-(\ref{eq:ge2})
are not time translation invariant, so that at which time one starts
is of relevance. The open set in the initial data will thus depend on
the starting time $\tau_{0}$. The condition demanded in this paper
is not only sufficient, but in fact also necessary for obtaining 
asymptotics of the form (\ref{eq:as1})-(\ref{eq:as2}) in the sense that,  
if a solution has this form of asymptotics, then for a late enough
time, the condition on the initial data will be satisfied. In this
sense, the condition described in this article is a characterization of
the solutions with asymptotic behaviour (\ref{eq:as1})-(\ref{eq:as2})
in terms of initial data. Observe finally that the condition, even
though it is formulated as a global condition on all of $S^{1}$ in
this paper, can be applied locally due to domain of dependence
arguments. Thus the condition prescribed here should be of relevance,
and should in fact be applicable in a neighbourhood of almost all
spatial points, even in the case with spikes. The problem in the
general case of course being that of proving that the evolution takes 
you to such a region. Finally, let us observe that this is not the
first result in this direction. In \cite{chr2}, Chru\'{s}ciel
considers developments of perturbations of initial
data for the Kasner $(\frac{2}{3},\frac{2}{3},-\frac{1}{3})$ metric
within the Gowdy class. He proves, among other things, curvature blow
up for these developments. In our setting the Kasner 
$(\frac{2}{3},\frac{2}{3},-\frac{1}{3})$ metrics correspond to 
$P=Q=0$.

\section{The equations and an outline of the argument}

Since the method does not depend on the dimension, and since the 
arguments in a sense become more transparent in a more general
setting, we will consider the following equations on $\mathbb{R}\times
T^{d}$:
\begin{eqnarray}
P\tt-\ett \Delta P-e^{2P}(Q_{\tau}^{2}-\ett |\nabla Q|^{2}) 
& = & 0 \label{eq:eq1} \\
P (\tau_{0},\cdot)\ =\  p_{0},\ \ \ 
P\t (\tau_{0},\cdot) & = & p_{1} \nonumber\\
Q\tt-\ett \Delta Q+2(P\t Q\t-\ett \nabla P\cdot \nabla Q) & = 
& 0 \label{eq:eq2} \\
Q (\tau_{0},\cdot)\ =\  q_{0},\ \ \
Q\t (\tau_{0},\cdot) & = & q_{1}. \nonumber
\end{eqnarray}
These equations have some similarities with wave maps. Let
\[
g=-dt\otimes dt+\sum_{i=1}^{d}dx^{i}\otimes dx^{i}
\]
be the Minkowski metric on $\mathbb{R}\times T^{d}$ and let
\[
g_{0}=dP\otimes dP+e^{2P}dQ\otimes dQ
\]
be a metric on $\mathbb{R}^{2}$. Observe that $(\mathbb{R}^{2},g_{0})$
is isometric to hyperbolic space. The wave map equations for a map 
from $(\mathbb{R}\times T^{d},g)$ to $(\mathbb{R}^{2},g_{0})$ is 
given by the Euler-Lagrange equations corresponding to the action
\[
\int
g_{0,ab}g^{\a\beta}\partial_{\alpha}u^{a}\partial_{\beta}u^{b}dtdx=
\int [-P_{t}^{2}+|\nabla P|^{2}-e^{2P}Q_{t}^{2}+e^{2P}|\nabla
Q|^{2}]dxdt.
\]
This should be compared with (\ref{eq:eq1}) and (\ref{eq:eq2}) which
are obtained as the Euler-Lagrange equations corresponding to the 
action
\[
\int [-P_{\tau}^{2}+e^{-2\tau}|\nabla P|^{2}-e^{2P}Q_{\tau}^{2}+
e^{2P-2\tau}|\nabla Q|^{2}]dxd\tau.
\]

The exact statement of the result can be found in section
\ref{section:conclusions}. It is however rather lengthy, and therefore
we wish to state a somewhat less technical consequence here. We need
to define some energy norms, but first note the following
convention concerning multi-indices. If
$\a=(\a_{1},...,\a_{d})$ where $\a_{i}$ are non-negative integers,
then
\[
\da f=\frac{\partial^{|\a|}f}{\partial x_{1}^{\a_{1}}\cdots 
\partial  x_{d}^{\a_{d}}},
\]
where $|\a|=\a_{1}+...+\a_{d}$. The natural energy norm for $P$ is
\begin{equation}\label{eq:vrekdef}
\vre_{k}(\tau)=\vre_{k}(P,\tau)=\frac{1}{2}\sum_{|\a|=k}\int_{T^{d}}
[(\da \dt P)^{2}+\ett |\nabla \da P|^{2}]d\theta
\end{equation}
and the one for $Q$ is
\begin{equation}\label{eq:ekdefp}
E_{k}(\tau)=E_{k}(P,Q,\tau)=
\frac{1}{2}\sum_{|\a|=k}\int_{T^{d}}[e^{2P}(\da\dt Q)^{2}+
e^{2P-2\tau}|\nabla\da Q|^{2}]d\theta.
\end{equation}
Finally it is natural to introduce
\[
\vrf_{k}(\tau)=\vrf_{k}(P,\tau)=\frac{1}{2}\sum_{|\a|=k}\int_{T^{d}}
|\nabla \da P|^{2}d\theta.
\]
If $p_{0},p_{1},q_{0},q_{1}\in C^{\infty}(T^{d},\mathbb{R})$, we will by
$\e_{k}(p_{0},p_{1},\tau)$ mean $\vre_{k}$ with $P$ replaced by $p_{0}$
and $P_{\tau}$ replaced by $p_{1}$. We associate
$e_{k}(p_{0},q_{0},q_{1},\tau)$ with $E_{k}$ and $\nu_{k}(p_{0})$ with
$\vrf_{k}$ similarly.

\begin{thm}
Let $p_{0},p_{1}\in C^{\infty}(T^{d},\mathbb{R})$ satisfy
\[
2\gamma\leq p_{1}\leq 1-2\gamma,
\]
where $\gamma>0$. Then, if $\tau_{0}$ is big enough and 
$e_{k}(p_{0},q_{0},q_{1},\tau_{0})$, $k=0,...,m_{d}=2[d/2]+2$ are 
small enough, we have smooth solutions to (\ref{eq:eq1}) and
(\ref{eq:eq2}) on $[\tau_{0},\infty)\times T^{d}$ with the 
following properties: there are $v,w,q\in C^{\infty}(T^{d},\mathbb{R})$
with
\[
\gamma\leq v\leq 1-\gamma
\]
and polynomials $\pi_{1,k},\pi_{2,k}$ in $\tau-\tau_{0}$ for every
non-negative integer $k$ such that
\[
\|P-\rho\|_{C^{k}(T^{d},\mathbb{R})}\leq
\pi_{1,k}\exp[-2\gamma(\tau-\tau_{0})],
\]
\[
\sum_{|\a|\leq k}\|e^{2\rho}
[\da Q-\da q]\|_{C(T^{d},\mathbb{R})}\leq
\pi_{2,k},
\]
where $\rho=v\cdot (\tau-\tau_{0})+w$.
\end{thm}

\textit{Remark}. The sizes of $\tau_{0}$ and the 
$e_{k}(p_{0},q_{0},q_{1},\tau_{0})$ only depend on $\gamma$,
$\e_{k}(p_{0},p_{1},\tau_{0})$ and $\nu_{k}(p_{0})$, $k=0,...,m_{d}$.
The reader interested in more conclusions is referred to section
\ref{section:conclusions}.

Let us give an outline of the proof. Given a smooth function $P$,
we can view (\ref{eq:eq2}) as a linear equation for $Q$. The central
part of the argument is an analysis of this linear equation assuming
$P$ satisfies
\[
0<\g\leq P_{\tau}\leq 1-\g
\]
and $\mathcal{E}_{k}(P)$ bounded for $k=0,...,m_{d}=2[d/2]+2$.
Under these conditions on $P$ one can prove that $E_{k}$
satisfies
\begin{equation}\label{eq:structure}
E_{k}^{1/2}(\tau)\leq\mathcal{P}_{k}(\tau-\tau_{0})\exp[-\g(\tau-\tau_{0})]
\end{equation}
for $k=0,...,m_{d}$. Here $\mathcal{P}_{k}$ is a polynomial in 
$\tau-\tau_{0}$ whose coefficients depend on the values of
$E_{k}$, $\vre_{k}$ and $\vrf_{k}$ at $\tau=\tau_{0}$ for 
$k=0,...,m_{d}$. 
The polynomials $\mathcal{P}_{k}$ have one important property.
If one lets $E_{k}(\tau_{0})$ go to zero, then the coefficients of
the polynomial go to zero, assuming the other parameters are constant.
In consequence, under these conditions on $P$, one has very good 
control of $Q$. The idea is then to consider the following iteration:
\begin{eqnarray}
P\nc\tt-\ett \Delta P\n-e^{2P\nmo}(Q_{n,\tau}^{2}-\ett 
|\nabla Q_{n}|^{2}) 
& = & 0 \label{eq:it2} \\
Q\nc\tt-\ett \Delta Q\n+2(P\nmoc\t Q\nc\t-\ett \nabla P\nmo\cdot 
\nabla Q\n) & = 
& 0 \label{eq:it1} \\
P{}_{0,}\tt-\ett\Delta P_{0} & = & 0 \label{eq:it0} 
\end{eqnarray}
where
\[
P_{n}(\tau_{0},\cdot)=p_{0},\
Q_{n}(\tau_{0},\cdot)=q_{0},\
P_{n,\tau}(\tau_{0},\cdot)=p_{1},\
Q_{n,\tau}(\tau_{0},\cdot)=q_{1}.
\]
Observe that we only need to solve linear equations during the iteration
in that if $P\nmo$ is given, then we can compute $Q\n$ using
(\ref{eq:it1}), so that (\ref{eq:it2}) also becomes a linear equation.
One sets up an induction hypothesis on $P_{n}$ amounting to the 
statements 
\begin{equation}\label{eq:c12}
0<\g\leq P_{n,\tau}\leq 1-\g\ \ \ \mathrm{and}\ \ \
\vre_{k}(P_{n},\tau)\leq c_{k}<\infty,\ \ k=0,...,m_{d}.
\end{equation}
It is not too difficult finding conditions on the initial data of
$P$ ensuring that $P_{0}$ satisfies these conditions. By the above
observations concerning the linear equation (\ref{eq:it1}), one gets
very good control of the behaviour of $Q_{n+1}$ if (\ref{eq:c12}) 
holds. Inserting this information into
(\ref{eq:it2}), it turns out that the propagation of the inductive 
hypotheses essentially boils down to a smallness 
condition on $E_{k}(\tau_{0})$ $k=0,...,m_{d}$ due to the structure
(\ref{eq:structure}). In this way, we produce a sequence of iterates
obeying (\ref{eq:c12}). The arguments proving 
convergence turn out to be similar to the arguments proving the 
propagation of the inductive hypotheses, and a smallness condition
on $E_{k}(\tau_{0})$ ensures the desired behaviour. Thus one produces
a solution to (\ref{eq:eq1}) and (\ref{eq:eq2}) with certain
properties. Due to the fact that one knows the solution to have these
extra properties, one can show that it has the desired asymptotic
behaviour.

\section{Local existence}

Let us here state the local existence result we will need.
We are interested in equations of the form
\begin{eqnarray}
\Box \bP (t,x) & = & \bF (t,x,\bP,\bP_{t},\nabla\bP)\label{eq:eq}\\
\bP(t_{0},\cdot) & = & \bp_{0},\ \ 
\bP_{t}(t_{0},\cdot)=\bp_{1}. \nonumber
\end{eqnarray}

\begin{prop}\label{prop:local}
Consider the equation (\ref{eq:eq}) where $t_{0}\in \mathbb{R}_{-}$, 
$(t,x)\in \mathbb{R}_{-}\times T^{d}$,
$(\bp_{0},\bp_{1})\in H^{k+1}(T^{d},\mathbb{R}^{l})\times
H^{k}(T^{d},\mathbb{R}^{l})$, $\bF$ is a smooth function
and $k\geq m_{d}/2=[d/2]+1$. Here, $\mathbb{R}_{-}=(-\infty,0)$. 
Then there are $T_{1},T_{2}\in \mathbb{R}_{-}$ with
$T_{1}<t_{0}<T_{2}$ such that there is a unique solution solution of 
(\ref{eq:eq}) in
\begin{equation}\label{eq:rightspace}
C(I,H^{k+1}(T^{d},\mathbb{R}^{l}))\cap
C^{1}(I,H^{k}(T^{d},\mathbb{R}^{l}))
\end{equation}
where $I=[T_{1},T_{2}]$.
Let $T_{\ma}$ be the supremum of the times $T\in \mathbb{R}_{-}$ 
such that there is 
a unique solution to (\ref{eq:eq}) in (\ref{eq:rightspace}) for
$I=[t_{0},T]$ and define $T_{\mi}$ similarly. Then either $T_{\ma}=0$, or
\[
\sup_{t\in [t_{0},T_{\ma})}
\|\bP\|_{C^{1}([0,t]\times T^{d},\mathbb{R}^{l})}=\infty.
\]
The statement for $T_{\min}$ is similar. 
\end{prop}
\textit{Remark}. The fact that $T^{d}$ is compact makes some of the 
usual conditions on $\bF$ unnecessary. 

The proof uses estimates of the form (6.4.5)' of \cite{hoer}
adapted to the torus case. A similar result for $k\geq m_{d}$ 
only requires Sobolev embedding and is sufficient for our purposes.
We will later solve the non-linear problem by carrying out an
iteration, and it will then be of interest to solve equations of 
the form 
\begin{eqnarray*}
Q_{tt}-\Delta Q=G_{1}Q_{t}+G_{2}\cdot\nabla Q+G_{3}\\
Q(t_{0},\cdot)=q_{0} \ \ \
Q_{t}(t_{0},\cdot)=q_{1}
\end{eqnarray*}
where $G_{1},G_{2},G_{3}\in C^{\infty}(\mathbb{R}_{-}\times T^{d},
\mathbb{R})$ and the initial data are smooth. Observe that local as
well as global existence of smooth solutions to this problem is
assured by Proposition \ref{prop:local}. Observe finally that equations
of the form 
\[
\bP_{\tau\tau} (\tau,x)-e^{-2\tau}\Delta\bP (\tau,x)  =  
\bF (\tau,x,\bP,\bP_{\tau},\nabla\bP)
\]
on $\mathbb{R}\times T^{d}$ can be transformed to equations of the
form (\ref{eq:eq}) by the transformation $t(\tau)=-e^{-\tau}$.

\section{The polarized case}

It is always instructive to start with an easier subcase, and in 
this case we have the added incentive that the zeroth iterate of the 
iteration belongs to this subclass, so let us consider the polarized
case. Let $P$ solve
\begin{equation}\label{eq:pol}
P_{\tau\tau}-e^{-2\tau}\Delta P=0.
\end{equation}
\begin{prop}
Consider a smooth solution $P$ to the polarized equation (\ref{eq:pol}). 
Then there are
$v,w\in C^{\infty}(T^{d},\mathbb{R})$ and positive constants
$C_{k}\in\mathbb{R}$ for every non-negative integer $k$ such that
\begin{equation}\label{eq:polcon}
\|P_{\tau}-v\|_{C^{k}(T^{d},\mathbb{R})}+
\|P-v\tau-w\|_{C^{k}(T^{d},\mathbb{R})}\leq C_{k}(1+\tau)e^{-2\tau}
\end{equation}
for all $\tau\geq 0$.
\end{prop}
\textit{Proof}.
The energy $\vre_{k}$ defined by (\ref{eq:vrekdef}) satisfies
\begin{equation}\label{eq:dedt}
\frac{d\vre_{k}}{d\tau}\leq 0.
\end{equation}
By Sobolv embedding, we conclude that 
\[
\sum_{|\a|=k}[\|\da\dt P\|_{C(T^{d},\mathbb{R})}+e^{-\tau}
\|\nabla\da P\|_{C(T^{d},\mathbb{R})}]\leq C_{k}<\infty
\]
for all $k$. Inserting this information into (\ref{eq:pol}), we
conclude that 
\[
\sum_{|\a|=k}\|\da\dt^{2} P\|_{C(T^{d},\mathbb{R})}\leq C_{k}e^{-\tau}
\]
for all $k$ and $\a$. We conclude the existence of a function
$v\in C^{\infty}(T^{d},\mathbb{R})$ such that 
\[
\|P_{\tau}-v\|_{C^{k}(T^{d},\mathbb{R})}\leq C_{k}e^{-\tau},
\]
which in its turn proves the existence of a $w\in 
C^{\infty}(T^{d},\mathbb{R})$ such that 
\[
\|P-v\tau-w\|_{C^{k}(T^{d},\mathbb{R})}\leq C_{k}e^{-\tau}.
\]
One consequence of this is of course that 
\[
\|P\|_{C^{k}(T^{d},\mathbb{R})}\leq C_{k}(1+\tau)
\]
for $\tau\geq 0$. Inserting this in (\ref{eq:pol}) and going through
the same steps as above, one ends up with (\ref{eq:polcon}).
$\hfill\Box$

Observe that in the end, it turns out that $P$ and all its spatial 
derivatives do not grow faster than linearly. However, the natural
consequence of the boundedness of $\vre_{k}$ is that expressions of 
the form
\[
\frac{1}{2}\sum_{|\a|=k}\int_{T^{d}}
\ett |\nabla \da P|^{2}d\theta
\]
are bounded. In other words, the form of the energy
(\ref{eq:vrekdef}), forced upon us by the energy methods, is not
well suited to the behaviour of the solutions. The way we achieved
the linear growth estimate in the proposition above, was through a 
procedure which was very wasteful of derivatives. This is not likely
to be successful in the general non-linear case. However, there is 
another point of view, and we will describe it in the next section.

\section{Energies}
In this section we gather some observations concerning the type
of energies we will be using.
Let $\tau_{0}\in\mathbb{R}_{+}=[0,\infty)$, $I=[\tau_{0},\infty)$ 
and let $\psi\in C^{\infty}(\mathbb{R}\times T^{d},\mathbb{R})$. Let
\begin{equation}\label{eq:varfdef}
\vrf_{k}(\psi,\tau)=\frac{1}{2}\sum_{|\a|=k}\int_{T^{d}}
|\nabla \da \psi|^{2}d\theta.
\end{equation}

\begin{lemma}\label{lemma:sup}
Assume $\psi\in C^{\infty}(\mathbb{R}\times T^{d},\mathbb{R})$ satisfies
\[
\vre_{k}^{1/2}(\psi,\tau)\leq\e_{k}<\infty 
\]
for $\tau\in I$ and $k=0,...,m$, where $\vre_{k}$ is defined in
(\ref{eq:vrekdef}), $m\geq m_{d}/2$, $m_{d}=2[d/2]+2$
and the $\e_{k}$ are constants. Then, for $k\leq m-m_{d}/2$, 
\[
\sum_{|\a|=k}[\|\da\dt \psi(\tau,\cdot)\|_{C(T^{d},\mathbb{R})}+
e^{-\tau}\|\nabla \da\psi(\tau,\cdot)\|_{C(T^{d},\mathbb{R})}]\leq
C(\e_{k}+\e_{k+m_{d}/2})
\]
for $\tau\in I$, where the $\e_{k}$ may be omitted if $k>0$. Furthermore, 
\begin{equation}\label{eq:linest}
\vrf_{k}^{1/2}(\psi,\tau)\leq C[\vrf_{k}^{1/2}(\psi,\tau_{0})+
\e_{k+1}(\tau-\tau_{0})]
\end{equation}
if $k\leq m-1$, so that
\[
\sum_{|\a|=k}\|\nabla\da\psi\|_{C(T^{d},\mathbb{R})}\leq C[
\vrf_{k+m_{d}/2}^{1/2}(\psi,\tau_{0})+\e_{k+1+m_{d}/2}(\tau-\tau_{0})]
\]
if $k\leq m-m_{d}/2-1$.
\end{lemma}
\textit{Remark}. When we write $\|f\|_{C(T^{d},\mathbb{R})}$ for a vector
valued function $f$, we mean the sup norm of the Euclidean norm
of the function.

\textit{Proof}. The first inequality follows from Sobolev embedding,
as well as the third, given the second. The second is proved by
computing
\[
\frac{d\vrf_{k}}{d\tau}=
\sum_{|\a|=k}\int_{T^{d}}
\nabla \da \psi\cdot \nabla \da\dt \psi d\theta\leq
2C\vrf_{k}^{1/2}\vre_{k+1}^{1/2}.
\]
$\hfill\Box$

Observe that the main point of this lemma is the estimate
(\ref{eq:linest}). We get a linear growth estimate for
$\vrf_{k}^{1/2}(\psi,\tau)$ if we know that $\mathcal{E}_{k+1}(\psi,\tau)$
is bounded for $\tau\in I$. In other words, there is a price for
this sort of estimate, but we only have to pay one derivative.

As has already been mentioned, when considering (\ref{eq:ge2}),
the following energy will be of interest,
\begin{equation}\label{eq:ekdef}
E_{k}(\eta,\xi,\tau)=
\frac{1}{2}\sum_{|\a|=k}\int_{T^{d}}[e^{2\eta}(\da\dt \xi)^{2}+
e^{2\eta-2\tau}|\nabla\da\xi|^{2}]d\theta.
\end{equation}

\begin{lemma}\label{lemma:rec}
Let $\tau_{0}\in \mathbb{R}_{+}$ and $I=[\tau_{0},\infty)$.
Assume $\eta,\ \xi\in C^{\infty}(\mathbb{R}\times T^{d},\mathbb{R})$
and that
\[
0<\gamma\leq \eta_{\tau}\leq 1-\gamma <1
\]
on $I\times T^{d}$ where $\gamma$ is a constant. Then
\begin{equation}\label{eq:dekestmod}
\frac{dE_{k}(\eta,\xi)}{d\tau}
\leq
-2\gamma E_{k}+
\sum_{|\a|=k}\int_{T^{d}}f_{\a}(\eta,\xi)
\da\dt \xi d\theta,
\end{equation}
where
\begin{equation}\label{eq:fkdef}
f_{\a}(\eta,\xi)=
\dt(e^{2\eta}\da\dt \xi)-\nabla\cdot(e^{2\eta-2\tau}\nabla\da\xi).
\end{equation}
If $\a=\ah+e_{l}$, where $e_{l}$ is an element of $\mathbb{Z}^{d}$ 
whose $l$:th component is $1$ and whose remaining components are zero, 
we have the following recursion formula for $f_{\a}$:
\begin{equation}\label{eq:frec}
f_{\a}=
\d_{l} f_{\ah}-2\d_{l}\eta f_{\ah}-
2\d_{l}\dt\eta e^{2\eta}\dah\dt \xi
+2\nabla(\d_{l}\eta)\cdot ( e^{2\eta-2\tau}\nabla\dah\xi).
\end{equation}
\end{lemma}

\textit{Proof}. 
Estimate
\[
\frac{dE_{k}}{d\tau}=\sum_{|\a|=k}\int_{T^{d}}
[\dt(e^{2\eta}\da\dt \xi)\da\dt \xi-
\eta\t e^{2\eta}(\da\dt \xi)^{2}+
\]
\[
+(\eta\t-1)e^{2\eta-2\tau}|\nabla\da\xi|^{2}
+e^{2\eta-2\tau}(\nabla\da\xi)\cdot(\nabla\da\dt \xi)]d\theta\leq
\]
\[
\leq
-2\gamma E_{k}+
\sum_{|\a|=k}\int_{T^{d}}[\dt(e^{2\eta}\da\dt \xi)-\nabla\cdot
(e^{2\eta-2\tau}\nabla\da\xi)]
\da\dt \xi d\theta,
\]
and (\ref{eq:dekestmod}) follows. We have 
\[
f_{\a}=\dt(e^{2\eta}\da\dt \xi)-\nabla\cdot
(e^{2\eta-2\tau}\nabla\da\xi)=
\]
\[
=\dt\d_{l}(e^{2\eta}\dah\dt \xi)-\dt(2\d_{l}\eta e^{2\eta}\dah\dt \xi)-
\d_{l}\nabla\cdot (e^{2\eta-2\tau}\nabla\dah\xi)+
\]
\[
+\nabla\cdot(2\d_{l}\eta e^{2\eta-2\tau}\nabla\dah\xi)=
\d_{l} f_{\ah}-2\d_{l}\eta f_{\ah}-
2\d_{l}\dt\eta e^{2\eta}\dah\dt \xi+
\]
\[
+2\nabla(\d_{l}\eta)\cdot ( e^{2\eta-2\tau}\nabla\dah\xi),
\]
proving (\ref{eq:frec}). $\hfill\Box$

\section{Iteration}

Consider the iteration (\ref{eq:it2})-(\ref{eq:it1}). 
We will only be interested in the future evolution of solutions to the
corresponding non-linear partial differential equation, and we will
implicitly assume the time interval on which our estimates are valid to
be $I=[\tau_{0},\infty)$. Let
\[
\pt_{n}=P_{n}-P_{n-1},\ \vre_{n,k}=\vre_{k}(P_{n},\cdot),\
\tvre_{n,k}=\vre_{k}(\pt_{n},\cdot),
\]
where $\vre_{k}$ is defined by (\ref{eq:vrekdef}), and
\[
\qt_{n}=Q_{n}-Q_{n-1},\ E_{n,k}=E_{k}(P_{n-1},Q_{n},\cdot),\
\et_{n,k}=E_{k}(P_{n-1},\qt_{n},\cdot)
\]
where $E_{k}$ is defined in (\ref{eq:ekdef}).
Observe that these expressions are all independent of $n$ if
we evaluate them at $\tau_{0}$, that $\vre_{n,k}$ is defined
if $n\geq 0$, $\tvre_{n,k}$ and $E_{n,k}$ are defined if
$n\geq 1$ and $\et_{n,k}$ is defined if $n\geq 2$. 

{\bf Conditions and conventions concerning initial data}. {\em We only
consider initial data} $(p_{0},p_{1},q_{0},q_{1})$ {\em with the property
that there is a} $\gamma>0$ {\em such that }
\begin{equation}\label{eq:a01}
0<2\gamma\leq p_{1}(\theta)\leq 1-2\gamma<1\ \ \forall\ \theta\in T^{d}.
\end{equation}
\textit{Secondly}, $\e_{k}$, $\nu_{k}$ \textit{and} $e_{k}$ {\em will
be taken to be constants satisfying}
\begin{equation}\label{eq:vnuk}
\vre^{1/2}_{n,k}(\tau_{0})\leq \e_{k},\
\mathcal{F}_{k}^{1/2}(P_{n},\tau_{0})\leq \nu_{k},\
E^{1/2}_{n,k}(\tau_{0})\leq e_{k}
\end{equation}
\textit{for} $k=0,...,m_{d}$, $m_{d}=2[d/2]+2$, \textit{with} 
$\e_{k}$ \textit{and} $e_{k}$ \textit{positive}.
 
{\bf Induction hypothesis.} {\em We assume that }
\begin{equation}\label{eq:inda1}
\vre^{1/2}_{n,k}(\tau)\leq \e_{k}+1\ \forall \tau\in I
\end{equation}
{\em and that the following inequality is fulfilled on} 
$I\times T^{d}$:
\begin{equation}\label{eq:inda4}
|P\nc\t(\tau,\theta)-P\nc\t(\tau_{0},\theta)|\leq\Delta\gamma.
\end{equation}

Here $\Delta\gamma$ should be suitably small relative to $\gamma$,
but not too small. We will below assume $\Delta\gamma=\gamma/4$
to hold.

Observe that if (\ref{eq:inda4}) holds for $n$ and $m$, then
\begin{equation}\label{eq:inda5}
|P_{n,\tau}(\tau,\theta)-P_{m,\tau}(\tau,\theta)|\leq2\Delta\gamma.
\end{equation}
Observe also that (\ref{eq:inda4}) implies that
\begin{equation}\label{eq:inda6}
0<2\gamma-\Delta\gamma\leq P\nc\t\leq 1-(2\gamma-\Delta\gamma)<1
\end{equation}
on $I\times T^{d}$. In the course of the argument, we will give
inequalities involving $\gamma,\e_{k},\nu_{k}$ and $e_{k}$
introduced in (\ref{eq:a01}) and (\ref{eq:vnuk}) such that if
they are fulfilled, the induction hypothesis is propagated.
By imposing additional requirements, one can then prove
\begin{equation}\label{eq:inda3}
\tvre^{1/2}_{n+1,\m}\leq
\frac{1}{2}\tvre^{1/2}_{n,\m}
\end{equation}
where
\begin{equation}\label{eq:tvrensup}
\tvre^{1/2}_{n,\m}=\sup_{\tau\in I}\tvre^{1/2}_{n,0}+...+
\sup_{\tau\in I}\tvre^{1/2}_{n,m_{d}}.
\end{equation}
Thus one obtains convergence. Observe that it makes sense to speak of
the suprema once we have proven that (\ref{eq:inda1}) is valid for all
$n$. Let us note some consequences of the induction hypothesis.

\begin{lemma}\label{lemma:pcon}
Assume (\ref{eq:inda1}) is satisfied up to and including $n-1$, 
$k=0,...,m_{d}$ and that (\ref{eq:vnuk}) holds. Then
\begin{equation}\label{eq:pl2}
\sum_{|\a|=k}\|\nabla \da P_{n-1}\|_{L^{2} (T^{d},\mathbb{R})}\leq 
C[\nu_{k}+(\e_{k+1}+1)(\tau-\tau_{0})]
\end{equation}
\begin{equation}\label{eq:tpl2}
\sum_{|\a|=k}\|\nabla \da\pt_{n-1}\|_{L^{2} (T^{d},\mathbb{R})}\leq 
C\tvre_{n-1,\m}^{1/2}\cdot (\tau-\tau_{0})
\end{equation}
on $I$, for $0\leq k\leq m_{d}-1$
Finally, for $n\geq 2$,
\begin{equation}\label{eq:tpsup}
|\pt_{n-1}(\tau,\theta)|\leq C\tvre_{n-1,\m}^{1/2}
\cdot(\tau-\tau_{0})
\end{equation}
on $I$.
\end{lemma}

\textit{Proof}. The estimates (\ref{eq:pl2}) and (\ref{eq:tpl2}) follow
from Lemma \ref{lemma:sup}. In order to prove 
(\ref{eq:tpsup}), we estimate
\[
|P_{n-1}(\tau,\theta)-P_{n-2}(\tau,\theta)|=
|\int_{\tau_{0}}^{\tau}\pt_{n-1,\tau}(s,\theta)ds|\leq
C\tvre_{n-1,\m}^{1/2}(\tau-\tau_{0}),
\]
where we have used Lemma \ref{lemma:sup} in order to 
obtain the last inequality. $\hfill\Box$

\begin{lemma}\label{lemma:zero}
If (\ref{eq:vnuk}) holds, then (\ref{eq:inda1})
holds for $n=0$.
\end{lemma}
\textit{Proof}. See (\ref{eq:dedt}). $\hfill\Box$

\begin{lemma}\label{lemma:phhsup}
There are constants $c_{d}$ such that if (\ref{eq:vnuk}) and
(\ref{eq:inda1}) are fulfilled for $n-1$ and
\begin{equation}\label{eq:cond1}
(\e_{2}+1)e^{-\tau_{0}}\leq c_{1}\gamma,
\end{equation}
if $d=1$, and
\begin{equation}\label{eq:condd}
[\nu_{m_{d}/2+1}+(\epsilon_{m_{d}/2+2}+1)]e^{-2\tau_{0}}\leq
c_{d}\gamma
\end{equation}
if $d\geq 2$, then
\[
\int_{\tau_{0}}^{\infty}e^{-2\tau}|\Delta P_{n-1}(\tau,\cdot)|
d\tau\leq \Delta\gamma/2
\]
on $T^{d}$.
\end{lemma}

\textit{Remark}. The expression $(\nu_{2}+\e_{3})\exp(-2\tau_{0})$
could also be used as the left hand side of (\ref{eq:cond1}) 
if one is prepared to keep track of one more derivative, 
c.f. the higher dimensional case of the argument presented in
this paper.

\textit{Proof}.
We get a division into two cases depending on the dimension $d$.
If $d\geq 2$, then $m_{d}\geq 4$, so that $m_{d}/2+1\leq m_{d}-1$. 
In consequence, we can use the estimates of Lemma \ref{lemma:pcon}
to obtain
\[
\int_{\tau_{0}}^{\infty}e^{-2\tau}|\Delta P_{n-1}(\tau,\cdot)|
d\tau\leq C\int_{\tau_{0}}^{\infty}e^{-2\tau}
[\nu_{m_{d}/2+1}+(\epsilon_{m_{d}/2+2}+1)(\tau-\tau_{0})]d\tau\leq
\]
\[
\leq C
[\nu_{m_{d}/2+1}+
(\epsilon_{m_{d}/2+2}+1)]e^{-2\tau_{0}}.
\]
If $d=1$, we do not have enough control to ensure the linear 
growth of the third spatial derivative in the $L^{2}$-norm, and 
therefore have to resort to using our control on $\mathcal{E}_{n-1,2}$.
We get
\[
\| P_{n-1,\theta\theta}\|_{C(S^{1})}\leq
C'\| P_{n-1,\theta\theta\theta}\|_{L^{2}(S^{1})}
\leq C e^{\tau}(\e_{2}+1),
\]
where we have used (\ref{eq:inda1}). Thus
\[
\int_{\tau_{0}}^{\infty}e^{-2\tau}|P_{n-1,\theta\theta}(\tau,\cdot)|
d\tau\leq C e^{-\tau_{0}}(\e_{2}+1)
\]
and the lemma follows. $\hfill\Box$

\begin{lemma}\label{lemma:zstep}
If the conditions of Lemma \ref{lemma:phhsup} and (\ref{eq:vnuk})
hold, the inductive hypotheses (\ref{eq:inda1}) and (\ref{eq:inda4})
are satisfied for $n=0$.
\end{lemma}
\textit{Proof}.
The lemma follows by combining Lemma \ref{lemma:zero},
\ref{lemma:phhsup} and (\ref{eq:it0}). $\hfill\Box$

\section{The n:th step}

The first task is to estimate the behaviour of $E_{n,k}$. 
The point of the argument
is to demand that all the iterates are such that $P_{n,\tau}$
belongs to a region where $Q_{n,\tau}$, at least intuitively,
should decay to zero exponentially. Consequently, we hope to
achieve an exponential decay for the energies $E_{n,k}$.
This is in fact the case, but as the argument is constructed, the natural
estimate that appears is a polynomial times an exponentially
decaying factor. The polynomials that appear have an 
important property we wish to formalize.
\begin{definition}
Let $\mathcal{P}$ be a polynomial in $\tau-\tau_{0}$
depending on the $\e_{k}$,  $e_{k}$ and $\nu_{k}$ for $k=0,...,m_{d}$.
We say that $\mathcal{P}$ is a $Q$-\textit{dominated 
polynomial} if the coefficients of $\mathcal{P}$ 
are polynomial in $e_{k}$, $\e_{k}$ and $\nu_{k}$ and go to
zero when the $e_{k}$ go to zero while the other expressions
are fixed.
\end{definition}
\textit{Remark}. The $Q$-dominated polynomials we will consider will
always be independent of $n$.

\begin{lemma}\label{lemma:step1}
Assume that (\ref{eq:vnuk}), (\ref{eq:inda1}) and (\ref{eq:inda4}) are 
satisfied for $n-1$ and let $0\leq k\leq m_{d}$. Then
\begin{equation}\label{eq:enkest}
E_{n,k}^{1/2}(\tau)\leq \mathcal{P}_{k}(\tau-\tau_{0})
\exp[-\gamma(\tau-\tau_{0})]
\end{equation}
on $I$, where $\mathcal{P}_{k}$ is a $Q$-dominated
polynomial. Furthermore,
\begin{equation}\label{eq:kqsup}
\sum_{|\a|=k}\{\|e^{P_{n-1}}\da\dt Q_{n}\|_{C(T^{d},\mathbb{R})}+
\|e^{P_{n-1}-\tau}\nabla\da Q_{n}\|_{C(T^{d},\mathbb{R})}\}\leq
\end{equation}
\[
\leq\mathcal{Q}_{k+1}(\tau-\tau_{0})
\exp[-\gamma(\tau-\tau_{0})]
\]
for $k=0,...,m_{d}/2$, where the $\mathcal{Q}_{k}$ are $Q$-dominated
polynomials.
\end{lemma}

Let us make some preliminary observations. By Lemma \ref{lemma:rec}, 
we have
\begin{equation}\label{eq:denkest}
\frac{d E_{n,k}}{d\tau}\leq -2\gamma E_{n,k}+\sum_{|\a|=k}
\int_{T^{d}}f_{n,\a}\da\dt Q_{n}d\theta
\end{equation}
where
\[
f_{n,\a}=
\dt(e^{2P_{n-1}}\da\dt Q_{n})-\nabla\cdot(e^{2P_{n-1}-2\tau}
\nabla\da Q_{n}).
\]
If $\a=\ah+e_{l}$, where $e_{l}$ is an element of $\mathbb{Z}^{d}$ 
whose $l$:th component is $1$ and whose remaining components are zero, 
we have the following recursion formula for $f_{n,\a}$:
\begin{equation}\label{eq:recfn}
f_{n,\a}=
\d_{l} f_{n,\ah}-2\d_{l} P_{n-1} f_{n,\ah}-
2\d_{l}\dt P_{n-1} e^{2 P_{n-1}}\dah\dt  Q_{n}
+
\end{equation}
\[
+2\nabla(\d_{l} P_{n-1})\cdot ( e^{2 P_{n-1}-2\tau}\nabla\dah Q_{n}).
\]
\begin{lemma}\label{lemma:aux}
If 
\begin{equation}\label{eq:ass}
\|e^{-P_{n-1}}f_{n,\a}\|_{L^{2}(T^{d},\mathbb{R})}\leq \pi_{k}\exp[-\gamma(\tau-\tau_{0})]
\end{equation}
for all $\a$ such that $|\a|=k$, where $\pi_{k}$ is a $Q$-dominated
polynomial independent of $n$, 
then an inequality of the form (\ref{eq:enkest}) holds.
\end{lemma}
\textit{Proof}. By (\ref{eq:denkest}) and (\ref{eq:ass}) we have
\[
\frac{d E_{n,k}}{d\tau}\leq -2\gamma E_{n,k}+\sum_{|\a|=k}
\int_{T^{d}}f_{n,\a}\da\dt Q_{n}d\theta\leq
\]
\[
\leq-2\gamma E_{n,k}
+\sum_{|\a|=k}\|e^{-P_{n-1}}f_{n,\a}\|_{L^{2}(T^{d},\mathbb{R})}
\|e^{P_{n-1}}\da\dt Q_{n}\|_{L^{2}(T^{d},\mathbb{R})}\leq
\]
\[
\leq-2\gamma E_{n,k}+\sqrt{2}C\pi_{k}\exp[-\gamma(\tau-\tau_{0})]
E_{n,k}^{1/2}.
\]
Thus
\[
\frac{d}{d\tau}\{\exp[2\gamma(\tau-\tau_{0})]E_{n,k}\}\leq
\sqrt{2}C\pi_{k}\{\exp[2\gamma(\tau-\tau_{0})]E_{n,k}\}^{1/2}
\]
which can be integrated to
\[
\{\exp[2\gamma(\tau-\tau_{0})]E_{n,k}\}^{1/2}\leq
E_{n,k}^{1/2}(\tau_{0})+
\int_{\tau_{0}}^{\tau}2^{-1/2}C\pi_{k}(s-\tau_{0})ds,
\]
and the lemma follows. $\hfill\Box$

\textit{Proof of Lemma \ref{lemma:step1}}. Observe that
\[
f_{n,0}=\dt(e^{2P_{n-1}}\dt Q_{n})-\nabla\cdot(e^{2P_{n-1}-2\tau}
\nabla Q_{n})=
\]
\[
=e^{2P_{n-1}}[Q_{n,\tau\tau}+2P_{n-1,\tau}Q_{n,\tau}-
\ett \Delta Q_{n}-2\ett \nabla P_{n-1}\cdot \nabla Q_{n}]=0.
\]
Let us also observe that when the two first terms in (\ref{eq:recfn})
hit an expression involving $\exp(2P_{n-1})$, then the effect is
to differentiate the expression regarding the exponential expression
mentioned as a constant. Inductively we thus get the conclusion that
if $|\a|=k+1$, then $f_{n,\a}$ consists of terms of the form
\begin{equation}\label{eq:typ1}
C_{\ae,\at}e^{2P_{n-1}}\dae\dt P_{n-1}\dat\dt Q_{n}
\end{equation}
and
\begin{equation}\label{eq:typ2}
B_{\ae,\at}e^{2P_{n-1}-2\tau}\nabla\dae P_{n-1}\cdot\nabla\dat Q_{n}
\end{equation}
where $|\ae|\geq 1$ and $\ae+\at=\a$. We will have to
use different estimates for different $k$:s. 

\textbf{Zeroth order energy}. Observe that
\[
\frac{d E_{n,0}}{d\tau}\leq -2\gamma E_{n,0}
\]
so that
\[
E_{n,0}^{1/2}(\tau)\leq e_{0}\exp[-\gamma(\tau-\tau_{0})]
\]
for $\tau\geq \tau_{0}$. Thus (\ref{eq:enkest}) holds for
$k=0$, with $\mathcal{P}_{0}=e_{0}$.

\textbf{Intermediate order energies}. The condition
$1\leq k\leq m_{d}/2$ defines what we mean by intermediate energies.
We carry out an argument by induction. 
Assume that we have (\ref{eq:enkest}) up to and including $k$,
$0\leq k\leq m_{d}/2-1$. By Lemma \ref{lemma:aux}, we need to consider
\[
\|e^{-P_{n-1}}f_{n,\a}\|_{L^{2}(T^{d},\mathbb{R})}
\]
for $|\a|=k+1$. In order to deal with terms of the form 
(\ref{eq:typ1}), we need to estimate
\begin{equation}\label{eq:t1}
\|e^{P_{n-1}}\dae\dt P_{n-1}\dat\dt Q_{n}\|_{L^{2}(T^{d},\mathbb{R})}
\end{equation}
where $\alpha_{1}+\alpha_{2}=\a$ and $|\alpha_{1}|\geq 1$.
By Lemma \ref{lemma:sup} and the induction hypothesis on $n$ (\ref{eq:inda1})
we can take out $\dae\dt P_{n-1}$ in the sup norm. The remaining part
is bounded by $\sqrt{2}E_{n,|\alpha_{2}|}^{1/2}$. By the induction
hypothesis on $k$ and the fact that $|\alpha_{2}|\leq k$, we get 
the conclusion that 
\[
\|e^{P_{n-1}}\dae\dt P_{n-1}\dat\dt Q_{n}\|_{L^{2}(T^{d},\mathbb{R})}\leq
\pi_{k+1}\exp[-\gamma(\tau-\tau_{0})].
\]
The estimate for terms of the form (\ref{eq:typ2}) is similar, and
we can thus apply Lemma \ref{lemma:aux} in order to obtain
(\ref{eq:enkest}) for the intermediate energies.

\textbf{High order energies}. By high order we mean 
$m_{d}/2+1\leq k\leq m_{d}$.
Since we cannot assume to control the sup norm of the derivatives
of $P_{n-1}$ indefinitely, we need to change our method. 
Observe that if $|\a|\leq k\leq m_{d}/2$, then
\begin{equation}\label{eq:supest}
\|\da (e^{P_{n-1}}Q_{n,\tau})\|_{L^{2}(T^{d},\mathbb{R})}\leq
\pi_{\a}\sum_{j=0}^{k}E_{n,j}^{1/2}
\end{equation}
for some polynomial $\pi_{\a}$, c.f. Lemma \ref{lemma:pcon}. 
The argument for 
$\exp(P_{n-1}-\tau)\nabla Q_{n}$ is similar, and we get
\begin{equation}\label{eq:qsup}
\| e^{P_{n-1}}Q_{n,\tau}\|_{C(T^{d},\mathbb{R})}+
\| e^{P_{n-1}-\tau}\nabla Q_{n}\|_{C(T^{d},\mathbb{R})}\leq
\mathcal{Q}_{1}\exp[-\gamma(\tau-\tau_{0})]
\end{equation}
by Sobolev embedding, where $\mathcal{Q}_{1}$ is a $Q$-dominated 
polynomial. Let us assume that (\ref{eq:enkest}) is satisfied up
to and including $m_{d}/2+l\leq m_{d}-1$, and that 
(\ref{eq:kqsup}) is satisfied for $0\leq k\leq l$. For $l=0$ we know this
to be true.  We wish to prove an estimate 
of the form (\ref{eq:ass}) and as for the intermediate energies, we 
need to consider (\ref{eq:t1}) where $\alpha_{1}+\alpha_{2}=\a$, $|\a|=
m_{d}/2+l+1$ and 
$|\alpha_{1}|\geq 1$. If $|\alpha_{1}|\leq m_{d}/2$, we can take out
$\dae\dt P_{n-1}$ in the sup norm to obtain
\[
\|e^{P_{n-1}}\dae\dt P_{n-1}\dat\dt Q_{n}\|_{L^{2}(T^{d},\mathbb{R})}\leq
\sqrt{2}\|\dae\dt P_{n-1}\|_{C(T^{d},\mathbb{R})}E_{n,|\alpha_{2}|}^{1/2}
\]
which satisfies a bound as in (\ref{eq:ass}) by Lemma \ref{lemma:sup},
the induction hypotheses on $l$ and the fact that $|\alpha_{2}|\leq
m_{d}/2+l$. If $|\alpha_{2}|\leq l$ we can
take out $e^{P_{n-1}}\dat\dt Q_{n}$ in the sup norm in order to achieve
a similar bound using the induction hypothesis on $l$ and the
boundedness of $\vre_{n,|\a_{1}|}$. The argument for terms 
of the form (\ref{eq:typ2})
is similar. Since $|\alpha_{1}|>m_{d}/2$ and $|\alpha_{2}|>l$ cannot 
occur at the same time, we are done. We have thus proven (\ref{eq:enkest})
for $k=m_{d}/2+l+1$. We now need to prove that (\ref{eq:kqsup}) holds
for $k=l+1$. However, this can be proven in the same way as 
(\ref{eq:supest}); replace $Q_{n,\tau}$ in that inequality with 
$\da\dt Q_{n}$ for $|\a|\leq l+1$. $\hfill\Box$

Let us now turn to the problem of estimating $\mathcal{E}_{n,k}$.
\begin{lemma}
Assume (\ref{eq:vnuk}), (\ref{eq:inda1}) and (\ref{eq:inda4}) are 
fulfilled for $n-1$ and let $k=0,...,m_{d}$. Then
\begin{equation}\label{eq:vrenkest}
\mathcal{E}_{n,k}^{1/2}(\tau)\leq
\mathcal{E}_{n,k}^{1/2}(\tau_{0})+
\int_{\tau_{0}}^{\tau}\mathcal{V}_{k}(s-\tau_{0})\exp[-2\gamma(s-\tau_{0})]ds,
\end{equation}
where $\mathcal{V}_{k}$ is a $Q$-dominated polynomial.
\end{lemma}
\textit{Proof}. Observe that 
\[
\frac{d\mathcal{E}_{n,k}}{d\tau}\leq
\sqrt{2}\sum_{|\alpha|=k}\| \da
[e^{2P\nmo}(Q_{n,\tau}^{2}-\ett 
|\nabla Q_{n}|^{2})]\|_{L^{2}(T^{d},\mathbb{R})}\mathcal{E}_{n,k}^{1/2}.
\]
It is thus of interest to estimate 
\[
\da (e^{2P\nmo}Q_{n,\tau}^{2})=
\]
\[
=\sum_{j=0}^{k}\sum_{\beta_{1}+...+
\beta_{j+2}=\a}
C_{\beta_{1},...,\beta_{j+2}}
D^{\beta_{1}}P_{n-1}\cdots D^{\beta_{j}}P_{n-1}
e^{2P_{n-1}}D^{\beta_{j+1}}\dt Q_{n}D^{\beta_{j+2}}\dt Q_{n}
\]
in $L^{2}$-norm. Consider a term in the sum. Observe that at most one 
$|\beta_{l}|$ can be greater than $m_{d}/2$. Combining Lemma
\ref{lemma:pcon} and \ref{lemma:step1}, we conclude that 
\[
\|\da (e^{2P\nmo}Q_{n,\tau}^{2})\|_{L^{2}(T^{d},\mathbb{R})}\leq
\pi_{k}\exp[-2\gamma(\tau-\tau_{0})]
\]
where $\pi_{k}$ is a $Q$-dominated polynomial. The argument for 
\[
\|\da (e^{2P\nmo-2\tau}|\nabla Q_{n}|^{2})\|_{L^{2}(T^{d},\mathbb{R})}
\]
is similar, and we obtain
\[
\frac{d\mathcal{E}_{n,k}}{d\tau}\leq
2\mathcal{V}_{k}\exp[-2\gamma(\tau-\tau_{0})]\mathcal{E}_{n,k}^{1/2},
\]
which can be integrated to (\ref{eq:vrenkest}). $\hfill\Box$

We now wish to specify conditions that imply (\ref{eq:inda1}) and 
(\ref{eq:inda4}) for $n$.

\begin{lemma}\label{lemma:indclo}
There are constants $c_{k}$, and non-negative integers
$i_{k}$ and $j_{k}$ such that if 
\begin{equation}\label{eq:eib}
e_{k}\leq c_{k}
(1+\e_{0}+\e_{m_{d}}+\nu_{m_{d}})^{-i_{k}}\gamma^{j_{k}}
\end{equation}
for $k=0,...,m_{d}$, and the relevant condition
in Lemma \ref{lemma:phhsup} is satisfied, then (\ref{eq:inda1}) and
(\ref{eq:inda4}) hold for $n$ if they hold for $n-1$.
\end{lemma}

\textit{Proof}.
Consider (\ref{eq:vrenkest}) in order to prove that (\ref{eq:inda1})
holds for $n$. We have 
\[
\mathcal{V}_{k}(s-\tau_{0})=\sum_{j=1}^{l_{k}}
a_{k,j}(s-\tau_{0})^{j}
\]
where the $a_{k,j}$ are polynomials in $\e_{i}$, $e_{i}$ and $\nu_{i}$
$i=0,...,m_{d}$ (observe that we can consider the coefficients to be
polynomials in $\nu_{m_{d}}$, $\epsilon_{0}$, $\epsilon_{m_{d}}$ and
the $e_{i}$ if we wish). We thus get 
\[
\int_{\tau_{0}}^{\tau}\mathcal{V}_{k}(s-\tau_{0})
\exp[-2\gamma(s-\tau_{0})]ds=
\sum_{j=1}^{l_{k}}\int_{\tau_{0}}^{\tau}a_{k,j}(s-\tau_{0})^{j}
\exp[-2\gamma(s-\tau_{0})]ds\leq
\]
\[
\leq
\sum_{j=1}^{l_{k}}a_{k,j}\frac{j!}{2^{j+1}\gamma^{j+1}}.
\]
We wish this expression to be less than or equal to $1$, and since
$a_{k,j}$ is polynomial in $e_{i}$, $\e_{i}$ and $\nu_{i}$ with each
term containing at least one factor $e_{i}$, we conclude that (\ref{eq:inda1})
holds assuming that an inequality of the form (\ref{eq:eib}) holds.
By (\ref{eq:it2}) and Lemma
\ref{lemma:phhsup}, we have
\[
|P_{n,\tau}(\tau,\theta)-P_{n,\tau}(\tau_{0},\theta)|\leq
\int_{\tau_{0}}^{\tau}[e^{-2s}|\Delta P_{n}|+
e^{2P\nmo}(Q_{n,\tau}^{2}+e^{-2s} |\nabla Q_{n}|^2)]ds\leq
\]
\[
\leq \Delta\gamma/2+
2\int_{\tau_{0}}^{\tau}\mathcal{Q}_{1}^{2}
\exp[-2\gamma(s-\tau_{0})]ds,
\]
by Lemma \ref{lemma:phhsup}. The remaining statement of the lemma
follows. $\hfill\Box$

\section{Convergence}
Consider the difference between (\ref{eq:it1}) for $n+1$ and $n$.
We have
\[
\qt_{n+1,\tau\tau}-\ett 
\Delta\qt_{n+1}+2(P_{n,\tau} \qt_{n+1,\tau}-\ett \nabla P_{n}\cdot 
\nabla\qt_{n+1})=
\]
\[
=-2(\pt_{n,\tau} Q_{n,\tau}-\ett \nabla\pt_{n}\cdot \nabla Q_{n})
\]
\begin{eqnarray}
\qt\npo (\tau_{0},\cdot) & = & 0 \label{eq:diff1}\\
\qt\npoc\t (\tau_{0},\cdot) & = & 0 \nonumber.
\end{eqnarray}

Since (\ref{eq:inda4}) is fulfilled for all $n\geq 0$ and
(\ref{eq:a01}) holds, Lemma 
\ref{lemma:rec} yields
\begin{equation}\label{eq:dtenkest}
\frac{d\et_{n,k}}{d\tau}\leq
-2\gamma\et_{n,k}+\sum_{|\a|=k}\int_{T^{d}}\ft_{n,\a}\da\dt\qt\n d\theta ,
\end{equation}
where
\[
\ft_{n,\a}=
\dt(e^{2P_{n-1}}\da\dt \qt\n)-\nabla\cdot(e^{2P_{n-1}-2\tau}
\nabla\da\qt_{n}).
\]
We have
\begin{equation}\label{eq:fnutn}
\ft_{n+1,0}=
e^{2P\n}[\qt_{n+1,\tau\tau}-\ett \Delta \qt_{n+1}+
2(P_{n,\tau} \qt_{n+1,\tau}-\ett \nabla P_{n}\cdot \nabla\qt_{n+1})]=
\end{equation}
\[
=-2e^{2P\n}(\pt_{n,\tau} Q_{n,\tau}-\ett \nabla\pt_{n}\cdot\nabla Q_{n}).
\]
The recursion formula is the same as before: if we have $\a=\ah+e_{l}$,
then
\[
\ft_{n+1,\a}=
\partial_{l} \ft_{n+1,\ah}-2\partial_{l}P_{n} \ft_{n+1,\ah}-
\]
\[
-2\partial_{l}\dt P_{n} e^{2P\n}D^{\ah}\dt \qt_{n+1}+2 
e^{2P\n-2\tau}\nabla\partial_{l}P_{n}\cdot\nabla\dah\qt_{n+1}.
\]

\begin{lemma}\label{lemma:conv}
Assume that conditions as in Lemma \ref{lemma:indclo} are fulfilled,
so that (\ref{eq:inda1})-(\ref{eq:inda5}) are fulfilled for all $n\geq 0$.
Let $0\leq k\leq m_{d}$, then
\begin{equation}\label{eq:tenkest}
\et_{n+1,k}^{1/2}\leq \tvre_{n,\m}^{1/2}\vpt_{k}(\tau-\tau_{0})
\exp[-(\gamma-2\Delta\gamma)(\tau-\tau_{0})],
\end{equation}
where $\vpt_{k}$ is a $Q$-dominated polynomial and $\tvre_{n,\m}^{1/2}$
is given by (\ref{eq:tvrensup}). Furthermore
\begin{equation}\label{eq:tkqsup}
\sum_{|\a|=k}\{\|e^{P\n}\da\dt \qt_{n+1} \|_{C(T^{d},\mathbb{R})}+
\|e^{P\n-\tau}\nabla\da\qt_{n+1} \|_{C(T^{d},\mathbb{R})}\}
\leq 
\end{equation}
\[
\leq\tvre_{n,\m}^{1/2}
\vqt_{k+1}(\tau-\tau_{0})\exp[-(\gamma-2\Delta\gamma)(\tau-\tau_{0})]
\]
on $I$, for $k=0,...,m_{d}/2$, where $\vqt_{k}$ is a $Q$-dominated
polynomial. 
\end{lemma}

The proof of this statement is similar to the proof of Lemma
\ref{lemma:step1}. The proof of the following lemma is analogous
to the proof of Lemma \ref{lemma:aux}

\begin{lemma}\label{lemma:aux2}
If 
\begin{equation}\label{eq:ass2}
\|e^{-P_{n}}\ft_{n+1,\a}\|_{L^{2}(T^{d},\mathbb{R})}\leq 
\tvre_{n,\m}^{1/2}\pi_{k}\exp[-(\gamma-2\Delta\gamma)(\tau-\tau_{0})]
\end{equation}
for all $\a$ such that $|\a|=k$, where $\pi_{k}$ is a $Q$-dominated
polynomial independent of $n$, then an inequality of the form
(\ref{eq:tenkest}) holds.
\end{lemma}

\textit{Proof of Lemma \ref{lemma:conv}}. 
Observe that for $|\a|\geq 1$, 
\begin{equation}\label{eq:obs}
\ft_{n+1,\a}=\sum_{\a_{1}+\a_{2}=\a}A_{\a_{1},\a_{2}}
e^{2P_{n}}(D^{\a_{1}}\dt\pt_{n}D^{\a_{2}}\dt Q_{n}
-\ett\nabla D^{\a_{1}}\pt_{n}\cdot\nabla D^{\a_{2}}Q_{n})
+
\end{equation}
\[
+\sum_{\beta_{1}+\beta_{2}=\a,|\beta_{1}|\geq 1}
B_{\beta_{1},\beta_{2}}e^{2P_{n}}(D^{\beta_{1}}\dt P_{n}
D^{\beta_{2}}\dt\qt_{n+1}-\ett
\nabla D^{\beta_{1}}P_{n}\cdot \nabla D^{\beta_{2}}\qt_{n+1}).
\]
In order to be able to apply Lemma \ref{lemma:aux2}, we need to
estimate the left hand side of (\ref{eq:ass2}). By the above 
observation (\ref{eq:obs}), it is enough to estimate the expressions
\begin{equation}\label{eq:typa1}
\|e^{P_{n}}D^{\a_{1}}\dt\pt_{n}D^{\a_{2}}\dt Q_{n}
\|_{L^{2}(T^{d},\mathbb{R})}+
\|e^{P_{n}-2\tau}\nabla D^{\a_{1}}\pt_{n}\cdot\nabla
D^{\a_{2}}Q_{n}
\|_{L^{2}(T^{d},\mathbb{R})}
\end{equation}
and
\begin{equation}\label{eq:typa2}
\|e^{P_{n}}D^{\beta_{1}}\dt P_{n}
D^{\beta_{2}}\dt\qt_{n+1}\|_{L^{2}(T^{d},\mathbb{R})}+
\|e^{P_{n}-2\tau}
\nabla D^{\beta_{1}}P_{n}\cdot 
\nabla D^{\beta_{2}}\qt_{n+1}\|_{L^{2}(T^{d},\mathbb{R})}.
\end{equation}
Concerning expressions of the form (\ref{eq:typa1}), we 
need only apply Lemma \ref{lemma:sup}, Lemma \ref{lemma:step1},
(\ref{eq:inda5}) and the fact that one of $|\a_{1}|$ and $|\a_{2}|$
has to be less than or equal to $m_{d}/2$ in order to bound (\ref{eq:typa1})
by the right hand side of (\ref{eq:ass2}). Let us illustrate on the 
first term in (\ref{eq:typa1})
under the assumption that $|\a_{1}|\leq m_{d}/2$. We have
\[
\|e^{P_{n}}D^{\a_{1}}\dt\pt_{n}D^{\a_{2}}\dt Q_{n}
\|_{L^{2}(T^{d},\mathbb{R})}\leq
\]
\[
\leq\|e^{P_{n}-P_{n-1}}\|_{C(T^{d},\mathbb{R})}
\|D^{\a_{1}}\dt\pt_{n}\|_{C(T^{d},\mathbb{R})}
\|e^{P_{n-1}}D^{\a_{2}}\dt Q_{n}
\|_{L^{2}(T^{d},\mathbb{R})}\leq
\]
\[
\leq C\exp[2\Delta\gamma(\tau-\tau_{0})]\tvre_{n,\m}^{1/2}
\mathcal{P}_{|\a_{2}|}\exp[-\gamma(\tau-\tau_{0})].
\]
If $\a=0$, then only terms of the form (\ref{eq:typa1}) are of
relevance, so in that case, (\ref{eq:tenkest}) follows.

\textbf{Intermediate energies}.
Let us now prove (\ref{eq:tenkest}) for $0\leq |\a|\leq m_{d}/2$ by
induction on $|\a|$. All we need to do is to prove that (\ref{eq:typa2})
is bounded by an expression as in the right hand side of (\ref{eq:ass2}).
Assume (\ref{eq:tenkest}) holds up to and including $k\leq m_{d}/2-1$.
Since $k+1\leq m_{d}/2$ and $|\beta_{1}|\leq |\a|=k+1$, we can use 
Lemma \ref{lemma:sup} in order to extract $D^{\beta_{1}}\dt P_{n}$
and $e^{-\tau}\nabla D^{\beta_{1}}P_{n}$ in the sup norm.
Since $|\beta_{2}|\leq |\a|-1\leq k$, we can estimate what remains
using (\ref{eq:tenkest}) and the inductive hypothesis. By applying 
Lemma \ref{lemma:aux2} we have thus proven (\ref{eq:tenkest}) for 
$k\leq m_{d}/2$. 

\textbf{High energies}. The argument now proceeds as in the proof
of Lemma \ref{lemma:step1}. By Lemma \ref{lemma:pcon}, $\da P_{n}$
is bounded in the sup norm by a polynomial if $|\a|\leq m_{d}/2$, so that 
the result concerning the intermediate energies yields the conclusion 
that 
\[
\|\da(e^{P_{n}}\qt_{n+1,\tau})\|_{L^{2}(T^{d},\mathbb{R})}
\]
can be bounded by the right hand side of (\ref{eq:tkqsup}) if $|\a|\leq m_{d}/2$. 
A similar statement holds for $e^{P_{n}-\tau}\nabla\qt_{n+1}$, and
(\ref{eq:tkqsup}) follows for $k=0$ by Sobolev embedding. Assume now
that (\ref{eq:tenkest}) holds up to and including $k=m_{d}/2+l\leq m_{d}-1$ and that
(\ref{eq:tkqsup}) holds for $k$ up to and including $l$. For $l=0$, we 
know this to be true. Consider now (\ref{eq:typa2}) with 
$\beta_{1}+\beta_{2}=\a$, $|\beta_{1}|\geq 1$ and $|\a|=k+1=m_{d}/2+l+1$.
If $|\beta_{1}|\leq m_{d}/2$, we can take out $D^{\beta_{1}}\dt P_{n}$
and $e^{-\tau}\nabla D^{\beta_{1}}P_{n}$ in the sup norm by
Lemma \ref{lemma:sup} and then apply the induction hypothesis to what
remains since $|\beta_{2}|\leq |\a|-1\leq m_{d}/2+l$. If $|\beta_{2}|\leq l$,
we can take out $e^{P_{n}}D^{\beta_{2}}\dt\qt_{n+1}$ and 
$e^{P_{n}-\tau}\nabla D^{\beta_{2}}\qt_{n+1}$ in the sup norm,
using the inductive hypothesis. What remains is bounded by 
$\mathcal{E}_{n,|\beta_{1}|}^{1/2}$ and (\ref{eq:tenkest}) follows for
$m_{d}/2+l+1$ since one of the inequalities $|\beta_{1}|\leq m_{d}/2$ and 
$|\beta_{2}|\leq l$ must hold. In order to prove (\ref{eq:tkqsup}) for 
$l+1$, we proceed by Sobolev embedding as before. $\hfill\Box$

Finally, let us consider $\tvre_{n,k}$.

\begin{lemma}\label{lemma:finals}
Assume that conditions as in Lemma \ref{lemma:indclo} are fulfilled,
so that (\ref{eq:inda1})-(\ref{eq:inda5}) are fulfilled for all $n\geq 0$.
We have 
\begin{equation}\label{eq:tvrenkest}
\tvre_{n+1,k}^{1/2}\leq
\tvre_{n,\m}^{1/2}
\int_{\tau_{0}}^{\tau}
\tilde{\mathcal{V}}_{k}(s-\tau_{0})
\exp[-2(\gamma-2\Delta\gamma)(s-\tau_{0})]ds,
\end{equation}
on $I$, for $k=0,...,m_{d}$, where $\tilde{\mathcal{V}}_{k}$ is a $Q$-dominated 
polynomial.
\end{lemma}

We need some preliminaries.
Consider the difference of (\ref{eq:it2}) for $n+1$ and $n$. We have
\[
\pt_{n+1,\tau\tau}-\ett \Delta\pt_{n+1}=
e^{2P_{n}}(Q_{n+1,\tau}^{2}-\ett |\nabla Q_{n+1}|^{2})-
\]
\[
-e^{2P_{n-1}}(Q_{n,\tau}^{2}-\ett |\nabla Q_{n}|^{2})=
e^{2P_{n}}(Q_{n+1,\tau}\qt_{n+1,\tau}-\ett \nabla Q_{n+1}\cdot
\nabla\qt_{n+1})+
\]
\[
+e^{2P_{n}}(Q_{n,\tau}\qt_{n+1,\tau}-\ett \nabla Q_{n}\cdot
\nabla\qt_{n+1})+
(e^{2P_{n}}-e^{2P_{n-1}})(Q_{n,\tau}^{2}-\ett |\nabla Q_{n}|^{2})=
\]
\[
=g_{1}+g_{2}+g_{3}.
\]
Estimate
\begin{equation}\label{eq:dtvrenkest}
\frac{d \tvre_{n+1,k}}{d\tau}\leq
\sum_{|\a|=k}\int_{T^{d}}\da(g_{1}+g_{2}+g_{3})\da\dt \pt_{n+1}d\theta.
\end{equation}
\begin{lemma}
If 
\begin{equation}\label{eq:ass3}
\|\da(g_{1}+g_{2}+g_{3})\|_{L^{2}(T^{d},\mathbb{R})}\leq
\tvre_{n,\m}^{1/2}\pi_{|\a|}\exp[-2(\gamma-2\Delta\gamma)(\tau-\tau_{0})]
\end{equation}
for $|\a|\leq m_{d}$, where $\pi_{|\a|}$ is a $Q$-dominated polynomial,
then (\ref{eq:tvrenkest}) follows.
\end{lemma}
\textit{Proof}. By (\ref{eq:dtvrenkest}), 
\[
\frac{d \tvre_{n+1,k}}{d\tau}\leq
\sum_{|\a|=k}
\sqrt{2}\|\da(g_{1}+g_{2}+g_{3})\|_{L^{2}(T^{d},\mathbb{R})}
\tvre_{n+1,k}^{1/2}
\]
which can be integrated to (\ref{eq:tvrenkest}) given the assumptions
of the lemma. $\hfill\Box$

\textit{Proof of Lemma \ref{lemma:finals}}. Consider the contribution
of the first term in $g_{1}$ to $\da g_{1}$. If more than $m_{d}/2$ 
derivatives hit one of $P_{n}$, $Q_{n+1,\tau}$ or $\qt_{n+1,\tau}$,
then everything else can be taken out in the sup norm, and we obtain
an estimate of the form (\ref{eq:ass3}). In fact the estimate is a
bit better, but we will need the form (\ref{eq:ass3}) for the other 
terms. The argument for the  contribution of the second term in $g_{1}$
is similar. The expression $\da g_{2}$ can be controlled
by similar arguments, but  we loose one factor 
$\exp[-2\Delta\gamma(\tau-\tau_{0})]$ in decay due to the fact that we have 
to compensate that we in some situations
have $e^{P_{n}}$ were we would prefer to have $e^{P_{n-1}}$.

Consider
\[
\da[(e^{2P_{n}}-e^{2P_{n-1}})Q_{n,\tau}^{2}].
\]
This expression is a linear combination of the following types of terms
\[
(e^{2P_{n}}D^{\a_{1}}P_{n}\cdots D^{\a_{l}}P_{n}-
e^{2P_{n-1}}D^{\a_{1}}P_{n-1}\cdots D^{\a_{l}}P_{n-1})
D^{\a_{l+1}}\dt Q_{n}D^{\a_{l+2}}\dt Q_{n}
\]
where $\a_{1}+...+\a_{l+2}=\a$. These terms can in turn be written as a
linear combination of terms such as
\begin{equation}\label{eq:type1}
e^{2P_{n-1}}D^{\a_{1}}P_{n}\cdots D^{\a_{r}}\pt_{n}
\cdot\cdot\cdot D^{\a_{l}}P_{n-1}
D^{\a_{l+1}}\dt Q_{n}D^{\a_{l+2}}\dt Q_{n}
\end{equation}
and
\begin{equation}\label{eq:type2}
(e^{2P_{n}}-
e^{2P_{n-1}})D^{\a_{1}}P_{n}\cdot\cdot\cdot D^{\a_{l}}P_{n}
D^{\a_{l+1}}\dt Q_{n}D^{\a_{l+2}}\dt Q_{n}.
\end{equation}
Terms of the form (\ref{eq:type1}) can be handled using Lemma
\ref{lemma:pcon} and \ref{lemma:step1}, since at most one $|\a_{j}|$
can be bigger than $m_{d}/2$. For terms of type (\ref{eq:type2}), we
estimate
\[
|(e^{2P_{n}}-
e^{2P_{n-1}})D^{\a_{1}}P_{n}\cdot\cdot\cdot D^{\a_{l}}P_{n}
D^{\a_{l+1}}\dt Q_{n}D^{\a_{l+2}}\dt Q_{n}|\leq
\]
\[
\leq 2|P_{n}-P_{n-1}|\exp[4\Delta\gamma(\tau-\tau_{0})]
e^{2P_{n-1}}|D^{\a_{1}}P_{n}\cdot\cdot\cdot D^{\a_{l}}P_{n}
D^{\a_{l+1}}\dt Q_{n}D^{\a_{l+2}}\dt Q_{n}|
\]
which we can deal with, using (\ref{eq:tpsup}). The contribution from
the second term in $g_{3}$ can be estimated similarly. $\hfill\Box$

\section{Conclusions}\label{section:conclusions}

\begin{thm}\label{thm:conclusions}
Let $0<\gamma\leq 1/4$, $\tau_{0}\geq 0$ and define $m_{d}=2[d/2]+2$. 
There are constants
$c_{i,d}$, $i=1,2$, and integers $l_{i,d}$, $i=1,2$ depending
on the dimension such that the following holds.
If $d=1$, let $\e_{2}$ and $\tau_{0}$ be such that 
\begin{equation}\label{eq:condd1}
(1+\e_{2})e^{-\tau_{0}}\leq c_{1,1}\gamma.
\end{equation}
If $d>1$, let $\nu_{m_{d}/2+1}$, $\e_{m_{d}/2+2}$ and $\tau_{0}$ be
such that
\begin{equation}\label{eq:conddd}
[\nu_{m_{d}/2+1}+\epsilon_{m_{d}/2+2}+1]e^{-2\tau_{0}}\leq
c_{1,d}\gamma.
\end{equation}
Specify the remaining $\epsilon_{k}$ and $\nu_{k}$, $k=0,...,m_{d}$ 
freely. Assume furthermore that
\begin{equation}\label{eq:ekcond}
e_{k}\leq c_{2,d}
(1+\e_{0}+\e_{m_{d}}+\nu_{m_{d}})^{-l_{1,d}}\gamma^{l_{2,d}}
\end{equation}
for $k=0,...,m_{d}$.
Then every quadruple of functions $(p_{0},p_{1},q_{0},q_{1})$ satisfying
\begin{equation}\label{eq:pocond}
2\gamma\leq p_{1}\leq 1-2\gamma
\end{equation}
and (\ref{eq:vnuk}), yield upon solving
(\ref{eq:eq1}) and (\ref{eq:eq2})
smooth solutions on $[\tau_{0},\infty)$ with the following properties.
For all non-negative integers $k$, there are polynomials  
$\Xi_{i,k}$, $i=1,...,6$ in $\tau-\tau_{0}$, and $v,w,q,r\in 
C^{\infty}(T^{d},\mathbb{R})$, where
\begin{equation}\label{eq:vcon}
0<\gamma\leq v\leq 1-\gamma<1
\end{equation}
on $T^{d}$ such that
\begin{equation}\label{eq:velcon}
\| P_{\tau}- v\|_{C^{k}(T^{d},\mathbb{R})}\leq
\Xi_{1,k}\exp[-2\gamma(\tau-\tau_{0})],
\end{equation}
\begin{equation}\label{eq:pconv}
\|P-\rho\|_{C^{k}(T^{d},\mathbb{R})}\leq
\Xi_{2,k}\exp[-2\gamma(\tau-\tau_{0})],
\end{equation}
\begin{equation}\label{eq:dtqb}
\sum_{|\a|\leq k}\|e^{2\rho}
\da \dt Q\|_{C(T^{d},\mathbb{R})}\leq
\Xi_{3,k},
\end{equation}
\begin{equation}\label{eq:qconv}
\sum_{|\a|\leq k}\|e^{2\rho}
[\da Q-\da q]\|_{C(T^{d},\mathbb{R})}\leq
\Xi_{4,k},
\end{equation}
\begin{equation}\label{eq:xi5}
\|e^{2\rho}Q_{\tau}-r\|_{C^{k}(T^{d},\mathbb{R})}\leq 
\Xi_{5,k}\exp[-2\gamma(\tau-\tau_{0})],
\end{equation}
and
\begin{equation}\label{eq:xi6}
\| e^{2\rho}(Q-q)+\frac{r}{2v}\|_{C^{k}(T^{d},\mathbb{R})}\leq
\Xi_{6,k}\exp[-2\gamma(\tau-\tau_{0})]
\end{equation}
for all $\tau\in [\tau_{0},\infty)$, where $\rho=v\cdot(\tau-\tau_{0})+w$.
Finally, assume $v\in C^{\infty}(T^{d},\mathbb{R})$ satisfies 
(\ref{eq:vcon}), and that
\begin{equation}\label{eq:converse}
\sum_{|\a|\leq m_{d}+1}\|e^{P}
\da \dt Q\|_{C(T^{d},\mathbb{R})}+
\| P_{\tau}- v\|_{C^{m_{d}+1}(T^{d},\mathbb{R})}\leq Ce^{-\epsilon\tau}
\end{equation}
where $\epsilon>0$ and $P,Q\in 
C^{\infty}(\mathbb{R}\times T^{d},\mathbb{R})$. Let 
\[
e_{k}'(\tau)=E_{k}^{1/2}(P,Q,\tau),\ \epsilon_{k}'(\tau)=
\vre_{k}^{1/2}(P,\tau),\
\nu_{k}'(\tau)=\mathcal{F}_{k}^{1/2}(P,\tau)
\]
and $\gamma'=\gamma/4$. Then for $\tau_{0}'$ big enough,
(\ref{eq:condd1})-(\ref{eq:pocond}) will be satisfied with
$e_{k}$ replaced by $e_{k}'(\tau_{0}')$, $\tau_{0}$ replaced with
$\tau_{0}'$, $\gamma$ replaced with $\gamma'$ etc. 
\end{thm}
\textit{Remark}. The last part of the theorem is intended to 
illustrate the fact that there is in some sense an equivalence 
between the conditions (\ref{eq:condd1})-(\ref{eq:pocond}) and
the asymptotics (\ref{eq:velcon})-(\ref{eq:xi6}).

Let us briefly comment on the conditions before we turn to the 
proof. The idea of the argument is to see to it that $P_{\tau}$
is always bounded away from $0$ below and $1$ above. 
As can be seen by Lemma \ref{lemma:phhsup}, the conditions 
(\ref{eq:condd1}) and (\ref{eq:conddd}) are there to ensure that 
the term $e^{-2\tau}\Delta P$ in (\ref{eq:eq1}) does not push $P_{\tau}$
out of this interval. The condition (\ref{eq:ekcond}) is then there
to ensure that the remaining terms in (\ref{eq:eq1}) do not push us
away from the desired region.

\textit{Proof}. Observe first that (\ref{eq:inda1}) and
(\ref{eq:inda4}) hold for $n=0$ due to Lemma \ref{lemma:zstep}
and the assumption that (\ref{eq:condd1}) or (\ref{eq:conddd})
hold. By Lemma \ref{lemma:indclo}, an estimate of the form
(\ref{eq:ekcond}), together with estimates of the form 
(\ref{eq:condd1}) or (\ref{eq:conddd}) imply that (\ref{eq:inda1})
and (\ref{eq:inda4}) hold for all $n\geq 0$. By (\ref{eq:tvrenkest}),
an argument similar to the proof of Lemma \ref{lemma:indclo} yields
that a condition of the type (\ref{eq:ekcond}) implies
\begin{equation}\label{eq:conv}
\tvre_{n+1,\m}^{1/2}\leq \frac{1}{2}\tvre_{n,\m}^{1/2}.
\end{equation}
Let $T\in I$ and let us consider the convergence on 
$[\tau_{0},T]\times T^{d}$. By (\ref{eq:tpsup}), we conclude
that $P_{n}$ is a Cauchy sequence in sup norm. Observe that under 
these circumstances, factors of the type $e^{2P_{n}}$ and
$e^{-2\tau}$ are of no importance, since we are considering a 
finite time interval, and since the sequence $P_{n}$ is uniformly
bounded on this finite time interval. By (\ref{eq:tkqsup}),
(\ref{eq:conv}) and the equations, we conclude that 
\[
\da\dt^{l}P_{n}\ \ \ \mathrm{and}\ \ \
\da\dt^{l}Q_{n}
\]
are Cauchy sequences in $C([\tau_{0},T]\times T^{d},\mathbb{R})$ 
for $1\leq |\a|+l\leq m_{d}/2+1$. The convergence of $Q_{n}$ 
follows from the convergence of $Q_{n,\tau}$, the
finiteness of the time interval and the fact that the iterates
coincide at $\tau_{0}$. In particular, the iteration yields 
$C^{2}$ solutions to the equations for $\tau\in [\tau_{0},\infty)$. 
By Proposition \ref{prop:local}, the solutions will be smooth if the
initial values consist of smooth functions (the transformation
$t=-e^{-\tau}$ yields an equation of the right form). 
Furthermore, if we let
\[
E_{k}(\tau)=E_{k}(P,Q,\tau)
\]
and
\[
\mathcal{E}_{k}(\tau)=\mathcal{E}_{k}(P,\tau),
\]
then these expressions will satisfy the estimates (\ref{eq:enkest}),
(\ref{eq:kqsup}) and (\ref{eq:inda1}).
\begin{lemma}
There are polynomials $\mathcal{P}_{k}$ and $\mathcal{Q}_{k}$ and
constants $c_{k}$ such that 
\begin{equation}\label{eq:ekest}
E_{k}^{1/2}(\tau)\leq \mathcal{P}_{k}(\tau-\tau_{0})
\exp[-\gamma(\tau-\tau_{0})],
\end{equation}
\begin{equation}\label{eq:psupest}
\sum_{|\a|=k}\{\|e^{P}\da \dt Q\|_{C(T^{d},\mathbb{R})}+
\|e^{P-\tau}\nabla\da  Q\|_{C(T^{d},\mathbb{R})}\}\leq
\end{equation}
\[
\leq\mathcal{Q}_{k+1}(\tau-\tau_{0})\exp[-\gamma(\tau-\tau_{0})]
\]
and
\begin{equation}\label{eq:vrekest}
\mathcal{E}_{k}\leq c_{k}<\infty
\end{equation}
on $I$, for all non-negative integers $k$.
\end{lemma}
\textit{Proof}.  
By our construction we know the statement
concerning (\ref{eq:ekest}) and (\ref{eq:vrekest}) to be true for 
$k=0,...,m_{d}$ and the statement concerning (\ref{eq:psupest})
to be true for $k=0,...,m_{d}/2$. We want to prove
the statement of the lemma by an inductive argument. Assume
it to be true up to and including $k\geq m_{d}$ for (\ref{eq:ekest})
and (\ref{eq:vrekest}) and to $k-m_{d}/2$ for (\ref{eq:psupest}).
Let us introduce
\[
\mathcal{F}_{k}=\frac{1}{2}\sum_{|\a|=k}
\int_{T^{d}}|\nabla\da  P|^{2}d\theta
\]
and
\[
\mathcal{G}_{k}=\exp[-\gamma(\tau-\tau_{0})]
\mathcal{F}_{k}.
\]
Our primary goal is to prove the following two inequalities
\begin{equation}\label{eq:diffin1}
\frac{dE_{k+1}}{d\tau}\leq
-2\gamma E_{k+1}+[\mathcal{Y}_{k+1,1}+\mathcal{Y}_{k+1,2}\vre_{k+1}^{1/2}]
\exp[-\gamma(\tau-\tau_{0})]E_{k+1}^{1/2}
\end{equation}
and
\begin{equation}\label{eq:diffin2}
\frac{d\vre_{k+1}}{d\tau}\leq
[\mathcal{Z}_{k+1,1}+\mathcal{Z}_{k+1,2}E_{k+1}^{1/2}
+\mathcal{Z}_{k+1,3}\mathcal{G}_{k}^{1/2}]
\exp[-\gamma(\tau-\tau_{0})]\vre_{k+1}^{1/2}
\end{equation}
where $\mathcal{Y}_{k+1,1}$, $\mathcal{Y}_{k+1,2}$,
$\mathcal{Z}_{k+1,1}$, $\mathcal{Z}_{k+1,2}$ 
and $\mathcal{Z}_{k+1,3}$ are polynomials in $\tau-\tau_{0}$. 
Observe that we also have
\begin{equation}\label{eq:diffin3}
\frac{d\mathcal{G}_{k}}{d\tau}\leq
-\gamma\mathcal{G}_{k}+C_{k}\exp[-\gamma(\tau-\tau_{0})/2]
\mathcal{G}_{k}^{1/2}\mathcal{E}_{k+1}^{1/2}.
\end{equation}
We have
\begin{equation}\label{eq:dekest}
\frac{dE_{k+1}}{d\tau}
\leq
-2\gamma E_{k+1}+
\sum_{|\a|=k+1}\int_{T^{d}}f_{\a}
\da \dt Q d\theta,
\end{equation}
where
\[
f_{\a}=f_{\a}(P,Q)=
\dt(e^{2P}\da \dt Q)-\nabla\cdot(e^{2P-2\tau}\nabla\da  Q)
\]
and if $\a=\ah+e_{l}$,
\begin{equation}\label{eq:reck}
f_{\a}=\partial_{l} f_{\ah}-2\partial_{l}P f_{\ah}-
\end{equation}
\[
-2\partial_{l}\dt P e^{2P}D^{\ah}\dt Q+
2\nabla\partial_{l}P\cdot e^{2P-2\tau}\nabla D^{\ah}Q.
\]
Let us first prove (\ref{eq:diffin1}).
Considering (\ref{eq:diffin1}) and (\ref{eq:dekest}), it is enough
to prove the estimate
\begin{equation}\label{eq:auxest}
\|e^{-P}f_{\a}\|_{L^{2}(T^{d},\mathbb{R})}\leq
[\Pi_{k+1,1}+\Pi_{k+1,2}\vre_{k+1}^{1/2}]\exp[-\gamma(\tau-\tau_{0})]
\end{equation}
when $|\a|=k+1$, where $\Pi_{k+1,1}$ and  $\Pi_{k+1,2}$ are polynomials.
Since $f_{0}$ is zero, we inductively get the 
conclusion that $f_{\a}$ will contain two types of terms:
\begin{equation}\label{eq:ptyp}
C_{\a_{1},\a_{2}}e^{2P}(D^{\a_{1}}\dt P)(D^{\a_{2}}\dt Q)
\end{equation}
and
\begin{equation}\label{eq:pthyp}
B_{\a_{1},\a_{2}}e^{2P-2\tau}\nabla D^{\a_{1}} P\cdot 
\nabla D^{\a_{2}} Q
\end{equation}
where $|\a_{1}|\geq 1$ and $\a_{1}+\a_{2}=\a$.
Consider terms of type (\ref{eq:ptyp}). Considering (\ref{eq:auxest}),
we wish to estimate
\[
\|e^{P}(D^{\a_{1}}\dt P)(D^{\a_{2}}\dt Q)\|_{L^{2}(T^{d},\mathbb{R})}.
\]
If $|\a_{1}|\leq k-m_{d}/2$, then we can estimate $D^{\a_{1}}\dt P$
in the sup norm using Lemma \ref{lemma:sup} and the inductive hypothesis
concerning (\ref{eq:vrekest}). Since $|\a_{2}|\leq |\a|-1\leq k$, the inductive
hypothesis concerning (\ref{eq:ekest}) yields the conclusion that
\[
\|e^{P}(D^{\a_{1}}\dt P)(D^{\a_{2}}\dt Q)\|_{L^{2}(T^{d},\mathbb{R})}
\leq C_{k+1}\mathcal{P}_{|\a_{2}|}\exp[-\gamma(\tau-\tau_{0})].
\]
If $|\a_{2}|\leq k-m_{d}/2$, then we can take out $e^{P}D^{\a_{2}}\dt Q$
in the sup norm, using the inductive assumption concerning 
(\ref{eq:psupest}), in order to obtain
\[
\|e^{P}(D^{\a_{1}}\dt P)(D^{\a_{2}}\dt Q)\|_{L^{2}(T^{d},\mathbb{R})}
\leq\sqrt{2}\mathcal{Q}_{|\a_{2}|+1}\exp[-\gamma(\tau-\tau_{0})]
\mathcal{E}_{|\a_{1}|}^{1/2}.
\]
Regardless of whether $|\a_{1}|=k+1$ or not, we get an estimate that 
fits into (\ref{eq:auxest}). As we cannot have $|\a_{2}|>k-m_{d}/2$ and
$|\a_{1}|>k-m_{d}/2$ at the same time, we have dealt with terms of the 
form (\ref{eq:ptyp}). Terms of the form (\ref{eq:pthyp}) can be handled
similarly. Equation (\ref{eq:diffin1}) follows.
 
As far as (\ref{eq:diffin2}) is concerned, we have 
\begin{equation}\label{eq:dvrek}
\frac{d\vre_{k+1}}{d\tau}\leq
\sum_{|\a|=k+1}\int_{T^{d}}\da [e^{2P}(Q^{2}_{\tau}-\ett 
|\nabla Q|^{2})]
\da \dt P d\theta.
\end{equation}
Considering (\ref{eq:diffin2}), it is thus enough to prove
\begin{equation}\label{eq:auxest2}
\|\da [e^{2P}(Q^{2}_{\tau}-\ett 
|\nabla Q|^{2})]\|_{L^{2}(T^{d},\mathbb{R})}\leq
\end{equation}
\[
\leq [\Pi_{k+1,1}+\Pi_{k+1,2}E_{k+1}^{1/2}
+\Pi_{k+1,3}\mathcal{G}_{k}^{1/2}]
\exp[-\gamma(\tau-\tau_{0})]
\]
for $|\a|=k+1$ and some polynomials $\Pi_{k+1,1}$, $\Pi_{k+1,2}$ and
$\Pi_{k+1,3}$. Consider
\begin{equation}\label{eq:sum}
\da (e^{2P}Q^{2}_{\tau})=
\sum_{l=0}^{k+1}\sum_{\a_{1}+...+\a_{l+2}=\a}C_{\a_{1},...,\a_{l+2}}
e^{2P}D^{\a_{1}}P\cdots D^{\a_{l}}P
D^{\a_{l+1}}\dt QD^{\a_{l+2}}\dt Q.
\end{equation}
Observe that we can bound $D^{\beta}P$ by a polynomial in supremum 
norm if $|\beta|\leq k-m_{d}/2$ and in $L^{2}$ norm if 
$|\beta|\leq k$ using the induction hypothesis
concerning (\ref{eq:vrekest}) and Lemma \ref{lemma:sup}. We can also control
$e^{P}D^{\beta}\dt Q$ in the sup norm, using (\ref{eq:psupest})
and the inductive assumption, if $|\beta|\leq k-m_{d}/2$, and in the
$L^{2}$ norm if $|\beta|\leq k$. Consider a term in the sum (\ref{eq:sum}).
At most one $|\a_{i}|$ can be greater than $k-m_{d}/2$. If all $\a_{i}$
satisfy $|\a_{i}|\leq k$, then we get a bound
\[
\|e^{2P}D^{\a_{1}}P\cdots D^{\a_{l}}P
D^{\a_{l+1}}\dt QD^{\a_{l+2}}\dt Q\|_{L^{2}(T^{d},\mathbb{R})}\leq
\Pi_{k+1}\exp[-2\gamma(\tau-\tau_{0})]
\]
where $\Pi_{k+1}$ is a polynomial. If $|\a_{i}|=k+1$, and $i\leq l$, 
i.e. if $k+1$ derivatives hit one $P$, then
\[
\|e^{2P}D^{\a_{1}}P\cdots D^{\a_{l}}P
D^{\a_{l+1}}\dt QD^{\a_{l+2}}\dt Q\|_{L^{2}(T^{d},\mathbb{R})}\leq
\]
\[
\leq \Pi_{k+1}\exp[-2\gamma(\tau-\tau_{0})]\|\da  P\|_{L^{2}(T^{d},\mathbb{R})}\leq
\Pi_{k+1}'\exp[-3\gamma(\tau-\tau_{0})/2]\mathcal{G}_{k}^{1/2}.
\]
Finally, if one $|\a_{i}|=k+1$, and $i>l$, i.e. if all the derivatives
hit one $\dt Q$, then 
\[
\|e^{2P}D^{\a_{1}}P\cdots D^{\a_{l}}P
D^{\a_{l+1}}\dt QD^{\a_{l+2}}\dt Q\|_{L^{2}(T^{d},\mathbb{R})}\leq
\]
\[
\leq\Pi_{k+1}\exp[-\gamma(\tau-\tau_{0})]\|e^{P}\da \dt
Q\|_{L^{2}(T^{d},\mathbb{R})}
\leq \Pi_{k+1}'\exp[-\gamma(\tau-\tau_{0})]E_{k+1}^{1/2}.
\]
As the argument for $\da (e^{2P-2\tau}|\nabla P|^{2})$ is similar 
(\ref{eq:auxest2}), and thereby (\ref{eq:diffin2}), follows.

Let
\[
\mathcal{H}_{k}=\mathcal{E}_{k+1}+E_{k+1}+
\mathcal{G}_{k}.
\]
The estimates (\ref{eq:diffin1})-(\ref{eq:diffin3}) imply
\begin{equation}\label{eq:thebound}
\frac{d\mathcal{H}_{k}}{d\tau}\leq
\mathcal{W}_{k,1}\exp[-\gamma(\tau-\tau_{0})/2]\mathcal{H}_{k}^{1/2}+
\mathcal{W}_{k,2}\exp[-\gamma(\tau-\tau_{0})/2]\mathcal{H}_{k},
\end{equation}
where $\mathcal{W}_{k,1}$ and $\mathcal{W}_{k,2}$ are polynomials.
This inequality implies that $\mathcal{H}_{k}$ is bounded on $I$.
In fact, since $a^{1/2}\leq\frac{1}{2}(1+a)\leq 1+a$ for $a\geq 0$, 
(\ref{eq:thebound}) yields
\[
\frac{d(1+\mathcal{H}_{k})}{d\tau}\leq
\{\mathcal{W}_{k,1}\exp[-\gamma(\tau-\tau_{0})/2]+
\mathcal{W}_{k,2}\exp[-\gamma(\tau-\tau_{0})/2]\}(1+\mathcal{H}_{k})
\]
whence $1+\mathcal{H}_{k}$ is bounded.
Thus $\vre_{k+1}$ is bounded, yielding (\ref{eq:vrekest}) 
for $k+1$, which when inserted in (\ref{eq:diffin1}) implies (\ref{eq:ekest})
for $k+1$, which, together with the induction hypothesis
yields (\ref{eq:psupest}) for $k+1-m_{d}/2$ by Sobolev embedding. $\hfill\Box$

Observe that using (\ref{eq:vrekest}) and Lemma 
\ref{lemma:sup}, we can conclude that
\[
|\da  P|\leq \mathcal{R}_{\a}
\]
on $I\times T^{d}$ for all $\a$, where $\mathcal{R}_{\a}$ are 
first degree polynomials in $\tau-\tau_{0}$. Using this together
with (\ref{eq:eq1}) and (\ref{eq:psupest}), we conclude that
\[
|\da \dt^{2} P|\leq \Gamma_{\a}\exp[-2\gamma(\tau-\tau_{0})]
\]
on $I\times T^{d}$ for all $\a$, where $\Gamma_{\a}$ are polynomials 
in $\tau-\tau_{0}$. This implies that $\dt P$ converges in any
$C^{k}(T^{d},\mathbb{R})$
norm as $\tau\rightarrow \infty$ to a smooth function $v$ on $T^{d}$,
and that (\ref{eq:velcon}) holds. Equation (\ref{eq:velcon}) in its 
turn implies that
\[
P(\tau,\cdot)-v\cdot(\tau-\tau_{0})\rightarrow w
\]
in any $C^{k}(T^{d},\mathbb{R})$, and that we have (\ref{eq:pconv}).
Observe now that by (\ref{eq:eq2}) we have
\[
\dt(e^{2P}\dt Q)=\nabla\cdot(e^{2P-2\tau}\nabla Q).
\]
Thus
\[
|\dt\da (e^{2P}Q_{\tau})|=|\nabla\cdot\da(e^{2P-2\tau}\nabla Q)|\leq
\pi_{\a}\exp[-2\gamma(\tau-\tau_{0})]
\]
combining (\ref{eq:psupest}) with the fact that $P-\tau\leq
C-\g(\tau-\tau_{0})$.
Here and below, $\pi_{\a}$ will denote a polynomial in $\tau-\tau_{0}$.
This inequality can be integrated in order to yield the conclusion
that there is an $r\in C^{\infty}(T^{d},\mathbb{R})$ such that
\begin{equation}\label{eq:preliminary}
|\da (e^{2P}Q_{\tau}-r)|\leq \pi_{\a}\exp[-2\gamma(\tau-\tau_{0})].
\end{equation}
Let now
\[
\rho=v\cdot(\tau-\tau_{0})+ w,
\]
and observe that we have (\ref{eq:pconv}).
We would now like to replace $P$ in (\ref{eq:preliminary})
with $\rho$. Consider for that reason
\[
\da (e^{2\rho}Q_{\tau}-r)=\da [(e^{2\rho}-e^{2P})Q_{\tau}]+
\da (e^{2P}Q_{\tau}-r).
\]
The second term we can bound using (\ref{eq:preliminary}). Let us 
consider the first term,
\[
\da [(e^{2\rho}-e^{2P})Q_{\tau}]=
\sum_{\a_{1}+\a_{2}=\a}C_{\a_{1},\a_{2}}
D^{\a_{1}}(e^{2\rho-2P}-1)D^{\a_{2}}(e^{2P}Q_{\tau}).
\]
The second factor on the right hand side is bounded by a constant
for $\tau\geq\tau_{0}$ by the above, and so the factor of interest
is
\[
D^{\a_{1}}(e^{2\rho-2P}-1).
\]
Consider first the case $\a_{1}=0$. If $\tau$ is great enough, then
$|2\rho-2P|\leq 1$, so that 
\[
|e^{2\rho-2P}-1|=|\sum_{n=1}^{\infty}\frac{(2\rho-2P)^{n}}{n!}|\leq
2|\rho-P|\exp[|2\rho-2P|]\leq 2e|\rho-P|\leq 
\]
\[
\leq \pi_{0}\exp[-2\gamma(\tau-
\tau_{0})],
\]
where we have used (\ref{eq:pconv}) in the last inequality.
If $|\a_{1}|\geq 1$ we get similar estimates for less complicated reasons.
To sum up,
\[
|\da (e^{2\rho}Q_{\tau}-r)|\leq \pi_{\a}\exp[-2\gamma(\tau-\tau_{0})],
\]
proving (\ref{eq:xi5}). We conclude that (\ref{eq:dtqb}) holds, and
therefore there is a $q\in C^{\infty}(T^{d},\mathbb{R})$ such
that (\ref{eq:qconv}) holds. Let us compute
\[
e^{2\rho}(Q-q)=e^{2\rho}(-\int_{\tau}^{\infty}Q_{\tau}(s)ds)=
-\int_{\tau}^{\infty}e^{2\rho(\tau)}Q_{\tau}(s)ds=
\]
\[
=-\int_{\tau}^{\infty}e^{2\rho(\tau)-2\rho(s)}
(e^{2\rho(s)}Q_{\tau}(s)-r)ds-
\int_{\tau}^{\infty}e^{2\rho(\tau)-2\rho(s)}rds.
\]
Observe that $r$ is independent of $s$ and compute
\[
\int_{\tau}^{\infty}e^{2\rho(\tau)-2\rho(s)}ds=
\int_{\tau}^{\infty}e^{-2v\cdot (s-\tau)}ds=\frac{1}{2v}.
\]
We thus conclude that
\[
\da (e^{2\rho}(Q-q)+\frac{r}{2v})=
-\da \left(\int_{\tau}^{\infty}e^{2\rho(\tau)-2\rho(s)}
(e^{2\rho(s)}Q_{\tau}(s)-r)ds\right)=
\]
\[
=-\sum_{\a_{1}+\a_{2}=\a}^{k}C_{\a_{1},\a_{2}}
\left(\int_{\tau}^{\infty}\dae
[e^{2\rho(\tau)-2\rho(s)}]
\dat(e^{2\rho(s)}Q_{\tau}(s)-r)ds\right).
\]
In order to estimate this expression, consider
\[
|\int_{\tau}^{\infty}\dae
[e^{2\rho(\tau)-2\rho(s)}]
\dat(e^{2\rho(s)}Q_{\tau}(s)-r)ds|\leq
\]
\[
\leq\left(\int_{\tau}^{\infty}[\dae
e^{2\rho(\tau)-2\rho(s)}]^{2}ds\right)^{1/2}
\left(\int_{\tau}^{\infty}
[\dat(e^{2\rho(s)}Q_{\tau}(s)-r)]^{2}ds\right)^{1/2}
\]
The integrand in the first factor can be bounded by a polynomial
in $s-\tau$ multiplied by $\exp[-4v\cdot (s-\tau)]$. In consequence
the first factor is bounded by a constant. The integrand appearing
in the second factor can be bounded by 
\[
\pi_{\a_{2}}\exp[-4\gamma(s-\tau_{0})]
\]
where $\pi_{\a_{2}}$ is a polynomial in $s-\tau_{0}$. Adding up these
observations, we get the conclusion
\[
|\da(e^{2\rho}(Q-q)+\frac{r}{2v})|\leq
\pi_{\a}\exp[-2\gamma(\tau-\tau_{0})],
\]
proving (\ref{eq:xi6}). 

Let us now prove that solutions with the desired asymptotics satisfy
the initial conditions at late enough times. Due to (\ref{eq:converse})
there is a $w\in C^{m_{d}+1}(T^{d},\mathbb{R})$ such that 
\begin{equation}\label{eq:p}
\| P(\tau,\cdot)-v\cdot
(\tau-\tau_{0})-w\|_{C^{m_{d}+1}(T^{d},\mathbb{R})}\rightarrow 0
\end{equation}
as $\tau\rightarrow \infty$. As a consequence, 
the $\e_{k}'(\tau)$ are bounded for the entire future and the 
$\nu_{k}'(\tau)$ do not grow faster than linearly. Furthermore,
by (\ref{eq:p}) and (\ref{eq:converse}),
\[
\sum_{|\a|\leq m_{d}+1}|D^{\a}\dt Q|
\leq C\exp(-v\cdot \tau-\epsilon\tau)
\]
Thus there is a $q\in C^{m_{d}+1}(T^{d},\mathbb{R})$ such that 
\[
\sum_{|\a|\leq m_{d}+1}\|e^{P}D^{\a}(Q-q)\|_{C(T^{d},\mathbb{R})}
\leq Ce^{-\epsilon\tau}.
\]
Since $P\leq C+(1-\gamma)\tau$ and $\gamma>0$, we get the conclusion
that $e_{k}'(\tau)$ decays to zero exponentially in time. The last
statement of the theorem follows. $\hfill\Box$  

\section{Curvature blow up}

Let us make some observations concerning the geometry of 
the metric (\ref{eq:gowdy}) given the conclusions of the previous
section.

\begin{prop}
Consider a metric of the form (\ref{eq:gowdy}). Assuming $P$ and $Q$
have the asymptotic behaviour obtained as a conclusion in Theorem
\ref{thm:conclusions}, we have
\[
\lim_{\tau\rightarrow \infty}\inf_{\theta\in S^{1}}
|(R_{\alpha\beta\gamma\delta}
R^{\alpha\beta\gamma\delta})(\tau,\theta)|=\infty.
\]
\end{prop}

\textit{Proof}. We proceed as in \cite{kar1}. Consider the 
orthonormal basis given by 
\[
e_{0}=e^{\lambda/4+3\tau/4}\partial_{\tau},\ \
e_{1}=e^{\lambda/4-\tau/4}\partial_{\theta},\ \
e_{2}=e^{\tau/2-P/2}\partial_{\sigma},\ \
e_{3}=e^{\tau/2+P/2}(\partial_{\delta}-Q\partial_{\sigma}).
\]
There is a natural scaling given by 
\[
\phi=e^{-\lambda/4-3\tau/4}.
\]
Only few terms in $\phi^{2}R_{\alpha\beta\gamma\delta}$, where we
assume that the indexes are with respect to the orthonormal basis
mentioned above, are non-negligible. This makes the computation of 
the Kretschmann scalar manageable and in fact,
\[
\phi^{4}R_{\alpha\beta\gamma\delta}
R^{\alpha\beta\gamma\delta}
\]
converges to a smooth non-zero function on $S^{1}$. Since $\lambda$ equals
$v^{2}(\theta)\tau$ up to something bounded, we get the conclusion of 
the proposition. $\hfill\Box$

There is also an elementary proof of causal geodesic incompleteness in
our situation.

\begin{prop}
Consider an inextendible causal geodesic $\gamma:(s_{-},s_{+})
\rightarrow M$, where $M=\mathbb{R}\times T^{3}$ with a metric
of the form (\ref{eq:gowdy}) where $P$ and $Q$ have the asymptotic
behaviour
obtained as a conclusion in Theorem \ref{thm:conclusions}. Assume
furthermore that 
$<\gamma'(s),\partial_{\tau}|_{\gamma(s)}><0$. Then $\gamma$ is future
incomplete and 
\[
\lim_{s\rightarrow s_{+}-}
|(R_{\alpha\beta\gamma\delta}
R^{\alpha\beta\gamma\delta})[\gamma(s)]|=\infty.
\]
\end{prop}

\textit{Proof}. Let the basis $e_{\mu}$ be as in the proof of the 
previous proposition and define
\[
f_{0}=-<\gamma',e_{0}|_{\gamma}>,\ \
f_{k}=<\gamma',e_{k}|_{\gamma}>
\]
for $k=1,2,3$. Observe that $\sum f_{k}^{2}\leq f_{0}^{2}$ due to 
causality. Furthermore, if we let the $\tau$ component of $\gamma$
be denoted by $\gamma_{0}$, then $d\gamma_{0}/ds>0$. Thus, if
$\gamma_{0}$ is bounded from above, it converges to a finite value
as $s\rightarrow s_{+}-$. 
By the causality of the curve and compactness of the spatial slices,
we then get the conclusion that $\gamma$ converges. Thus $\gamma$ is
continuously extendible, leading to the conclusion that it is 
extendible considered as a geodesic. This contradicts our assumptions,
and we conclude that $\gamma_{0}(s)\rightarrow \infty$ as 
$s\rightarrow s_{+}-$. Consider
\[
\frac{df_{0}}{ds}=-<\gamma',\nabla_{\gamma'}e_{0}>=
-\sum_{\mu,\nu}f_{\mu}f_{\nu}<e_{\mu},\nabla_{e_{\nu}}e_{0}>
\circ\gamma .
\]
Let $\phi$ be as in the previous proposition. 
Using the assumptions concerning the asymptotics one sees that 
\[
\phi<e_{1},\nabla_{e_{1}}e_{0}>=-\frac{1}{4}(\lambda_{\tau}-1),\ \
\phi<e_{2},\nabla_{e_{2}}e_{0}>=-\frac{1}{2}(1-P_{\tau})
\]
and
\[
\phi<e_{3},\nabla_{e_{3}}e_{0}>=-\frac{1}{2}(1+P_{\tau})
\]
and that all other elements of the matrix 
$\phi<e_{\mu},\nabla_{e_{\nu}}e_{0}>$ converge to zero exponentially
with $\tau$. Letting
\[
\theta_{k}=\phi\circ\gamma <e_{k},\nabla_{e_{k}}e_{0}>\circ\gamma ,
\]
we have
\[
\frac{df_{0}}{ds}=-\psi\circ\gamma \sum_{k}f_{k}^{2}
\theta_{k}+\psi\circ\gamma \delta_{1}f_{0}^{2}\geq
\]
\[
\geq\psi\circ\gamma f_{0}^{2}\frac{1}{4}(\lambda_{\tau}\circ\gamma -1)
+\psi\circ\gamma \delta_{1}f_{0}^{2},
\]
if $s$ is close enough to $s_{+}$, where $\psi=1/\phi$ and
$\delta_{1}(s)\rightarrow 0$ as $s\rightarrow s_{+}-$ (observe that
$\theta_{1}>0$ and $\theta_{2},\theta_{3}<0$ if $s$ is close enough to
$s_{+}$). Compute
\[
\frac{d\psi\circ\gamma}{ds}=\frac{\partial\psi}{\partial\tau}
\circ\gamma \frac{d\gamma_{0}}{ds}+
\frac{\partial\psi}{\partial\theta}
\circ\gamma \frac{d\gamma_{1}}{ds}
\]
where $\gamma_{1}$ is the $\theta$-coordinate of $\gamma$ (observe that
even though this is not well defined, the derivative is). However,
\[
\frac{d\gamma_{0}}{ds}=\psi\circ\gamma f_{0},\ \
\frac{d\gamma_{1}}{ds}=\exp[\lambda\circ\gamma /4-\gamma_{0}/4]
f_{1}
\]
so that 
\[
\frac{d\psi\circ\gamma}{ds}=
\psi^{2}\circ\gamma f_{0}[\frac{1}{4}(\lambda_{\tau}\circ\gamma +3)+
\delta_{2}]
\]
where $\delta_{2}(s)\rightarrow 0$ as $s\rightarrow s_{+}-$. Letting
$g=f_{0}\cdot \psi\circ\gamma$, we get 
\[
\frac{dg}{ds}=\frac{df_{0}}{ds}\psi\circ\gamma+
f_{0}\frac{d\psi\circ\gamma}{ds}\geq g^{2}[\frac{1}{4}
(\lambda_{\tau}\circ\gamma-1)
+\delta_{1}+\frac{1}{4}(\lambda_{\tau}\circ\gamma+3)+\delta_{2}].
\]
Thus there is an $s_{1}$ such that for $s\geq s_{1}$
\[
\frac{dg}{ds}(s)\geq \frac{1}{2}g^{2}(s),
\]
since $\lambda_{\tau}$ is bounded from below by a positive constant
for large $\tau$. We thus get the conclusion that the geodesic is
future incomplete. Since $\gamma_{0}(s)\rightarrow \infty$ as 
$s\rightarrow s_{+}-$, the statement concerning curvature blow up
follows from the previous proposition. $\hfill\Box$

\section{Conclusions and observations}

This paper provides a proof of the statement that an open set of 
initial data yields asymptotics of the form (\ref{eq:as1}) and
(\ref{eq:as2}). The fact that we have a condition on initial data 
will hopefully be useful when trying to make further
progress in understanding Gowdy spacetimes. However, the method of
proof relies heavily on the fact that the non-linear terms in the equations
die out exponentially. This will not be the case in general and 
different methods will have to be developed. Some intuition for what 
should happen has been developed in \cite{bag}, but how to make these
ideas rigorous is far from clear. Since approaching the 
general problem is difficult, it is natural to try to find an easier
problem which can be used as a model for some of the dynamics. One
way of obtaining a model problem is to carry out intuitive arguments
similar to those in \cite{bag}. For example, if $P_{\tau}>1$, 
$e^{P}Q_{\tau}$ should be negligible and $Q_{\theta}$ should be 
a time independent function. Inserting these assumptions in
(\ref{eq:ge1}) one obtains
\[
P_{\tau\tau}-e^{-2\tau}P_{\theta\theta}=-e^{2P-2\tau}Q_{\theta}^{2}.
\]
Replacing $Q_{\theta}$ with $1$  and calling $P-\tau$
$P$, one gets the equation
\begin{equation}\label{eq:model}
P_{\tau\tau}-e^{-2\tau}P_{\theta\theta}=-e^{2P}.
\end{equation}
This equation would, in the terminology of \cite{bag}, model a bounce. 
Considering an arbitrary solution to (\ref{eq:model}), $P_{\tau}$
should converge exponentially to a smooth negative function on
$S^{1}$. One question of interest would then be to find out how
fast this process of convergence occurs. This is an example of a 
model problem which is certainly easier to handle, but which still
contains some of the dynamics of the real situation. Since most
of the results concerning the asymptotic behaviour of Einstein's 
equations in the spatially inhomogeneous case make assumptions that
exclude the possibility of the non-linear terms being of any
importance, the study of even a simple model problem where this is
not the case should be of interest.

\section*{Acknowledgements}

Part of this work was carried out while the author was enjoying the
hospitality of The Erwin Schr\"{o}dinger International Institute for
Mathematical Physics.

\end{document}